\documentclass[a4paper]{article}
\usepackage{graphicx}
\usepackage{amsmath}
\usepackage{amssymb}

\newcommand{\beq}[1][]{\begin{equation}\label{#1}}
\newcommand{\eeq}{\end{equation}}

\newcommand{\om}{\ensuremath{\omega}}
\newcommand{\kap}{\ensuremath{\varkappa}}
\newcommand{\bxi}{\ensuremath{\boldsymbol{\xi}}}
\newcommand{\A}{\ensuremath{\boldsymbol{A}}}
\newcommand{\B}{\ensuremath{\boldsymbol{B}}}
\newcommand{\f}{\ensuremath{\boldsymbol{f}}}

\title{$q$-Breathers in Fermi-Pasta-Ulam chains: 
existence, localization and stability}

\author{S.~Flach$^1$, M.~V.~Ivanchenko$^2$ and O.~I.~Kanakov$^2$
\\
\\
$^1$: Max-Planck-Institut f\"ur Physik komplexer Systeme,
\\
N\"othnitzer Str. 38, D-01187 Dresden, Germany
\\
$^2$: Department of Radiophysics, Nizhny Novgorod
University,
\\
Gagarin Avenue, 23, 603950 Nizhny Novgorod, Russia
}

\begin{document}

\maketitle

\begin{abstract}
The Fermi-Pasta-Ulam (FPU) problem consists of
the nonequipartition of energy among normal modes of a weakly
anharmonic atomic chain model. In the harmonic limit each normal
mode corresponds to a periodic orbit in phase space and is
characterized by its wave number $q$. We continue normal modes
from the harmonic limit into the FPU parameter regime and obtain
persistence of these periodic orbits, termed here $q$-Breathers (QB).
They are characterized by time periodicity, exponential
localization in the $q$-space of normal modes and linear
stability up to a size-dependent threshold amplitude. 
Trajectories computed in the original FPU setting are
perturbations around these exact QB solutions. The QB concept
is applicable to other nonlinear lattices as well.
\end{abstract}

\section{Introduction}

In the year 1955 E.~Fermi, J.~Pasta, and S.~Ulam (FPU) published
their celebrated report on thermalization of arrays of
particles connected by weakly nonlinear springs \cite{fpu}.
Instead of the expected equipartition of energy among 
the normal modes of the systems,
FPU observed 
that energy, initially
placed in a low frequency normal mode of the linear problem with
a frequency $\omega_q$ and a corresponding wave number $q$,
stayed almost completely locked within a few neighbour modes,
instead of being distributed among all modes of the system.
Moreover, recurrence of energy to the originally excited mode
was observed. The door was thus opened to study the 
fundamental physical and mathematical problem
of energy equipartition and ergodicity in nonlinear systems,
which involves the Kolmogorov-Arnold-Moser (KAM) theorem,
dividing thresholds between regular and chaotic dynamics,
and soliton bearing integrable models.

From the present perspective, the FPU observation (equally
coined FPU problem or FPU paradox) appears to consist of three major
ingredients: (FPU-1) for suitable parameter ranges (energy, system size,
nonlinearity strength) low frequency excitations are localized
in $q$-space of the normal modes; (FPU-2) recurrence of energy to
an initially excited low-frequency mode is observed; (FPU-3)
different thresholds upon tuning the parameters are observed -
a weak stochasticity threshold (WST) which separates regular from
chaotic dynamics, yet possibly preserving the localization character
in $q$-space, and an equipartition threshold (ET) also coined strong stochasticity
threshold separating localized from delocalized dynamics in $q$-space
(see \cite{Ford, chaosfpu, Izrailev} for a review).

Two major approaches were developed. The first one,
taken by N.~Zabusky and M.~Kruskal, was to analyze dynamics of the nonlinear
chain in the continuum limit, which led to a pioneering observation of
solitary waves \cite{Zabusky}. 
It took (FPU-1) as given, and aimed at obtaining quantitative estimates for (FPU-2).
A second approach was proposed by F. Izrailev and
B. Chirikov \cite{Izrailev_Chirikov} who associated energy equipartition with
dynamical chaos and aimed at an analytical estimate of the ET 
by computing the overlap of nonlinear resonances \cite{Chirikov} which leads
to strong dynamical chaos insuring energy equipartition. Below the threshold
the dynamics is regular and, thus, no equipartition should occur.
It aimed mainly at (FPU-3). 
Several other analytical \cite{deLuca, Shepel} and numerical 
\cite{italian},\cite{kantz},\cite{lcmcsmpegdc97}
threshold estimates have been published since, and will be discussed below.
Note that similar effects have been observed in many other nonlinear discrete chain
or field equations on a finite spatial domain, see e.g. \cite{pbasbbal70}.

In this work we show that stable periodic orbits, coined $q$-breathers (QB),
persist in the nonlinear FPU chain, which are exponentially localized in $q$-space.
The existence of these orbits was first reported in ref.\cite{sfmvioik05}.
Stability of these periodic orbits implies that nearby trajectories will
be localized in $q$-space as well, and evolve in a nearly regular fashion 
for long times on submanifolds similar to low-dimensional
tori in phase space. Recurrence times - to come close to an initial point on 
such a
torus again - will be much shorter than the general recurrence times
estimates for the dynamics on an assumed KAM torus in the whole phase space. 
That follows from the fact that a KAM torus in general will have the dimension
equivalent to the number of degrees of freedom, and thus much larger than the
effective dimension of the perturbed $q$-breather evolution.
Upon tuning
the parameters of the system, QBs will turn from stable to unstable
at certain threshold values, allowing for low-dimensional
chaotic evolution of nearby trajectories. The localization length in $q$-space
depends on these parameters as well, and at critical parameter values QBs delocalize,
leading to equipartition. The $q$-breather concept allows thus to address
simultaneously all three FPU ingredients. At the same time it can be extended
to completely different lattice systems and even to nonlinear field equations.

In the present work we construct QBs continuing them from the
linear case and study their properties both numerically and by
an analytical asymptotic calculation. We compare the thresholds
of QB existence and stability to the various stochasticity
thresholds mentioned above. Finally we show the persistence of
QBs in thermal equilibrium and during long transient processes.

\section{The model}
The FPU system is a chain of $N$ equal masses coupled by
nonlinear springs with the equations of motion containing
quadratic (the $\alpha$-model)
\begin{equation}
\label{eq1}
 \ddot{x}_n=(x_{n+1}-2x_n+x_{n-1})\\
+\alpha[(x_{n+1}-x_n)^2-(x_n-x_{n-1})^2]
\end{equation}
or cubic (the $\beta$-model)
\begin{equation}
\label{eq2}
 \ddot{x}_n=(x_{n+1}-2x_n+x_{n-1})
+\beta[(x_{n+1}-x_n)^3-(x_n-x_{n-1})^3]
\end{equation}
interaction terms, where $x_n$ is the displacement of the $n$-th
particle from its original position, and fixed boundary
conditions are taken $x_0=x_{N+1}=0$. A canonical transformation
\beq
x_n(t)=\sqrt{\frac{2}{N+1}}\sum\limits_{q=1}^N
Q_q(t)\sin{\left(\frac{\pi q n}{N+1}\right)}
\eeq
 takes into the reciprocal wavenumber space with $N$ normal mode coordinates
$Q_q(t)$. The equations of motion then read
\begin{equation}
\label{alpmod}
 \Ddot{Q}_q+\omega_q^2 Q_q=-\frac{\alpha}{\sqrt{2(N+1)}}\sum\limits_{l,m=1}^N
 \omega_q\omega_l\omega_m B_{q,l,m} Q_l Q_m
\end{equation}
for the FPU-$\alpha$ chain (\ref{eq1}) and
\begin{equation}
\label{betmod}
 \Ddot{Q}_q+\omega_q^2 Q_q=-\frac{\beta}{2(N+1)}\sum\limits_{l,m,n=1}^N
 \omega_q\omega_l\omega_m\om_n C_{q,l,m,n} Q_lQ_mQ_n
\end{equation}
for the FPU-$\beta$ chain (\ref{eq2}).
Here the coupling coefficients $B_{q,l,m}$ and $C_{q,l,m,n}$ are given by
\beq[Bqlm]
B_{q,l,m}=\sum\limits_{\pm}(\delta_{q\pm l \pm m,0}-\delta_{q\pm l \pm m,2(N+1)})\mbox{,}
\eeq
\beq[Cqlmn]
C_{q,l,m,n}=\sum\limits_{\pm}
(\delta_{q\pm l \pm m \pm n,0}-\delta_{q\pm l \pm m \pm n,2(N+1)}
-\delta_{q\pm l \pm m \pm n,-2(N+1)})
\mbox{.}
\eeq
The sum in \eqref{Bqlm} and \eqref{Cqlmn} is taken over all 4 or 8
combinations of signs, respectively. The normal mode frequencies 
\beq[omegas]
\omega_q=2\sin\frac{\pi q}{2(N+1)}
\mbox{.}
\eeq
are nondegenerate.

\section{A crosslink to discrete breathers}

The system of equations (\ref{alpmod}),(\ref{betmod}) corresponds to
a network of oscillators with different eigenfrequencies. These oscillators
are interacting with each other via {\sl nonlinear} interaction terms,
yet being long-ranged in $q$-space. Let us discuss the relation
of this problem to the wellknown field of discrete breathers \cite{DBs,DBs2,sfag05}.

Neglecting the 
nonlinear terms in the equations of motion
the $q$-oscillators get decoupled, each conserving its energy
\beq[Eq]
E_q=\frac{1}{2}\left(\dot{Q}_q^2+\omega_q^2 Q_q^2\right)
\eeq
in time. Especially, we may consider the excitation of only one of the
$q$-oscillators, i.e. $E_{q} \neq 0$ for $q\equiv q_0$ only. Such excitations
are trivial time-periodic and $q$-localized solutions (QBs) for
$\alpha=\beta=0$. 

This setting is similar to the case of
{\it discrete breathers} (DB), that are time-periodic and
spatially localized excitations e.g. on networks of interacting
{\sl identical anharmonic} oscillators, which survive continuation
from the trivial limit of zero coupling \cite{mackayaubry}.
Notably, DBs exist also in FPU lattices \cite{sfag05} and existence proofs
have been obtained as well \cite{lsm97,akk01,gj01}.
The reason for the generic existence of DBs is two-fold:
the nonlinearity of each oscillator allows to tune its excitation
frequency out of resonance with other nonexcited oscillators.
The case of a {\sl linear} coupling on the lattice ensures
a bound spectrum of small amplitude plane waves, and thus allows for the escape
of resonances of a DB frequency and its higher harmonics with that spectrum.
The spatial localization of DBs in such a case is typically 
exponential for short-range interactions \cite{DBs}.
Among the wealth of theoretical results we stress two here.
First, if the coupling on the lattice is short-ranged but purely
{\sl nonlinear}, the DB localizes in space {\sl superexponentially} \cite{compactons}.
Second, if the coupling is algebraically decaying on the lattice,
DBs localize only algebraically as well \cite{longrange}.

Let us compare these findings to the $q$-breather setting.
The {\sl nonidentical} $q$-oscillators are harmonic, but with
frequencies being nonresonant. That ensures continuation of
a trivial QB orbit from zero nonlinearity into the 
domain of nonzero nonlinearity.
Indeed, similar to the existence proof of discrete breathers
by MacKay and Aubry \cite{mackayaubry} by continuation from
the trivial uncoupled limit, we can start at zero nonlinearity
and excite mode $q_0$ to energy $E_{q_0}$, with all other modes
at rest. By properly choosing the origin of time the solution
will be time-periodic and time-reversible, i.e. 
$\hat{Q}_q(t+T)=\hat{Q}_q(t)$,
$\hat{Q}_q(t)=\hat{Q}_q(-t)$. Following
\cite{mackayaubry} we first observe that 
\begin{equation}
u_q(t)=\Ddot{U}_q+\omega_q^2 U_q + 
\frac{\alpha}{\sqrt{2(N+1)}}\sum\limits_{l,m=1}^N
\omega_q\omega_l\omega_m B_{q,l,m} U_l U_m
\label{functionmap}
\end{equation} 
maps any choice of time-periodic time-reversible functions $U_q(t;T)$
onto time-periodic and time-reversible functions $u_q(t;T)$
where the period $T$ is variable.
For $\alpha=0$ the particular choice $U_q(t;T)=\hat{Q}_q(t)$ 
and $T=2\pi/\omega_{q_0}$ ensures
that $u_q(t;T)=0$, i.e. the map (\ref{functionmap}) of time-periodic
and time-reversible functions (spanning a Banach space)
contains a zero. 
In the next step we have to check whether 
the map (\ref{functionmap}) is invertible
at $\alpha=0$. Following MacKay and Aubry \cite{mackayaubry} 
invertibility is trivially obeyed for all $q\neq q_0$
in (\ref{functionmap}), provided
$m 2\pi/T \neq \omega_q$. This nonresonance condition
is generically fulfilled for any finite lattice size $N$
and a period $T=2\pi/\omega_{q_0} +\epsilon$ where
$\epsilon$ is a small but finite number of the order $1/N$.
As for $q_0$ itself, we conclude
that invertibility is given provided we demand e.g.
fixing
the value $U_{q_0}(t=0;T)$ to a given nonzero number.
With these possibilities we arrive at the full invertibility
of (\ref{functionmap}) at $\alpha=0$. We may then conclude
(and skip the full derivation which follows \cite{mackayaubry})
that the Implicit Function Theorem can be applied, and
the zero of (\ref{functionmap}) for $\alpha=0$ will
persist up to some nonzero value of $\alpha$.
Of course the same procedure can be applied to the FPU-$\beta$ model.

As for their degree of localization in $q$-space (if any),
the above results on DBs suggest that two mechanisms counteract:
purely nonlinear interaction favours stronger than exponential
localization, while long-range interaction tends to delocalize
the QB. We also recall that the seemingly simple problem of
periodic motion of a classical particle in an anharmonic potential
of the FPU type follows from the differential equation
\begin{equation}
\ddot{x} = -x -\alpha x^2 - \beta x^3\;.
\label{aho}
\end{equation}
Bounded motion at some energy $E$ 
yields a solution which is periodic with some period $T(E)=2\pi/\Omega$
and can be represented by a Fourier series 
\begin{equation}
x(t) = \sum_k A_k {\rm e}^{ik\Omega t}
\label{ahosolution}
\end{equation}
which leads to algebraic equations for the Fourier coefficients
$A_k$
\begin{equation}
A_k = k^2 \Omega^2 A_k - \alpha \sum_{k_1} A_{k_1} A_{k-k_1}
-\beta \sum_{k_1,k_2} A_{k_1} A_{k_2} A_{k-k_1-k_2}\;.
\label{ahofc}
\end{equation}
Note that the equations (\ref{ahofc}) have similar properties
as compared to (\ref{alpmod}),(\ref{betmod}) - interaction between
the Fourier coefficients is nonlinear but long ranged.
Yet it is well known that the bounded solutions (\ref{ahosolution})
to (\ref{aho}) are analytic functions $x(t)$ and thus the
Fourier series coefficients $A_k$ converge exponentially fast with
$k$ \cite{zygmund1968}.

Let us summarize this chapter with stating that from
a mathematical point of view $q$-breather existence and their properties
can be treated in a similar way to the methodology of discrete breather
theory. The two different types of excitations, localized in real space
and localized in reciprocal $q$-space, have much in common
when represented in their natural phase space basis choice.
Both representations are connected to each other by a simple
canonical transformation, which is nothing but a rotation 
of the phase space basis. 
In the following we will use analytical and computational
tools developed for discrete breathers and analyze the properties
of $q$-breathers. 

\section{q-Breathers in the $\alpha$-FPU system}

\subsection{Numerical results}

\begin{figure}[t]
{
\centering
\resizebox*{0.90\columnwidth}{!}{\includegraphics{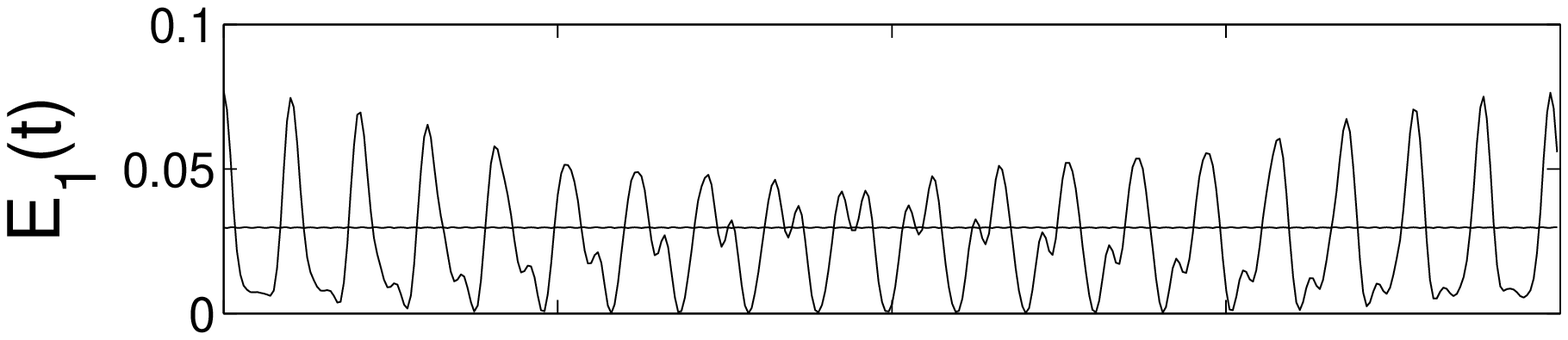}}
\resizebox*{0.90\columnwidth}{!}{\includegraphics{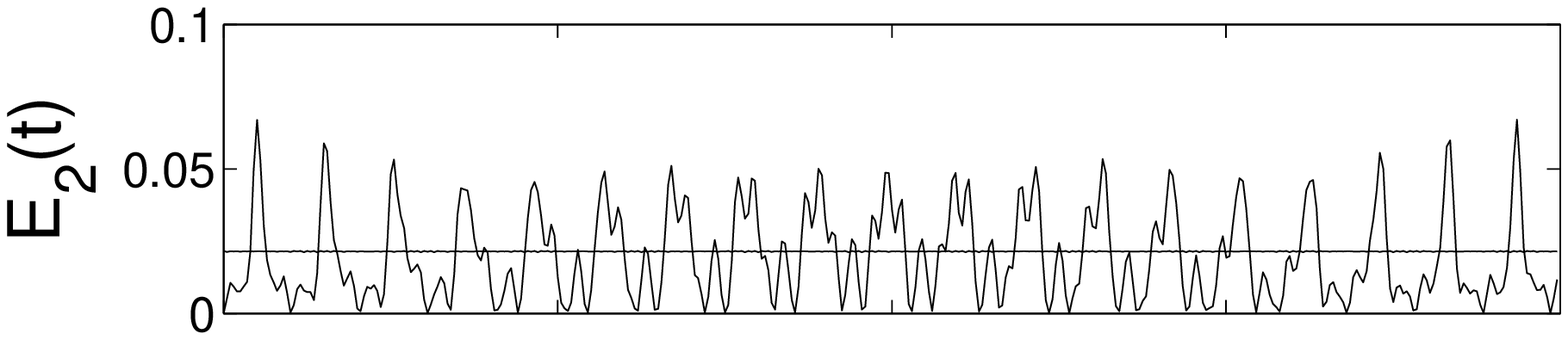}}
\resizebox*{0.90\columnwidth}{!}{\includegraphics{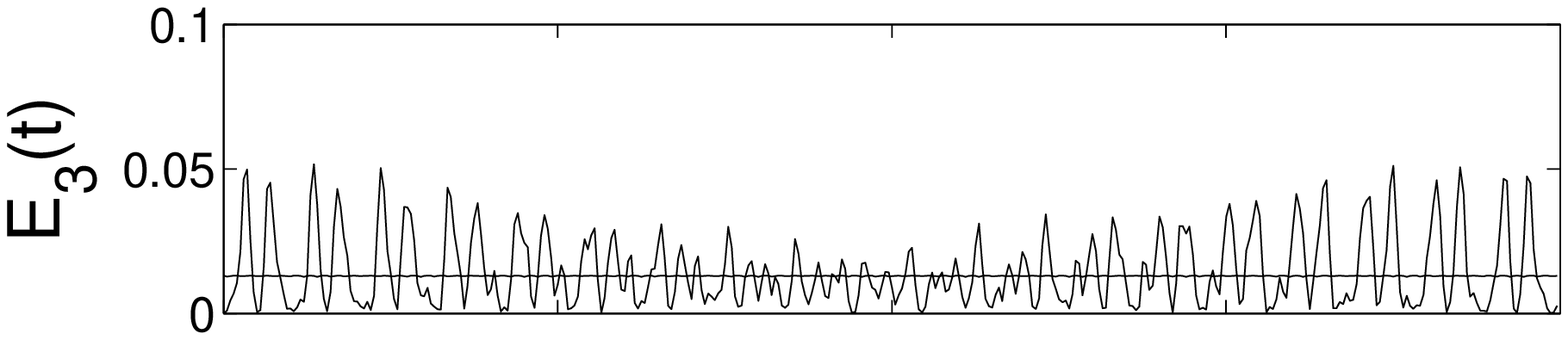}}
\resizebox*{0.90\columnwidth}{!}{\includegraphics{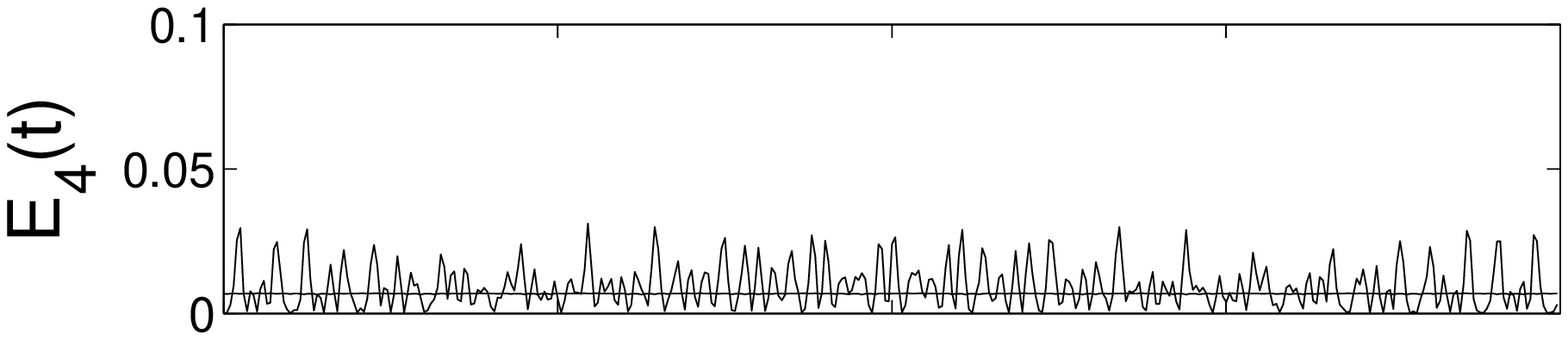}}
\resizebox*{0.90\columnwidth}{!}{\includegraphics{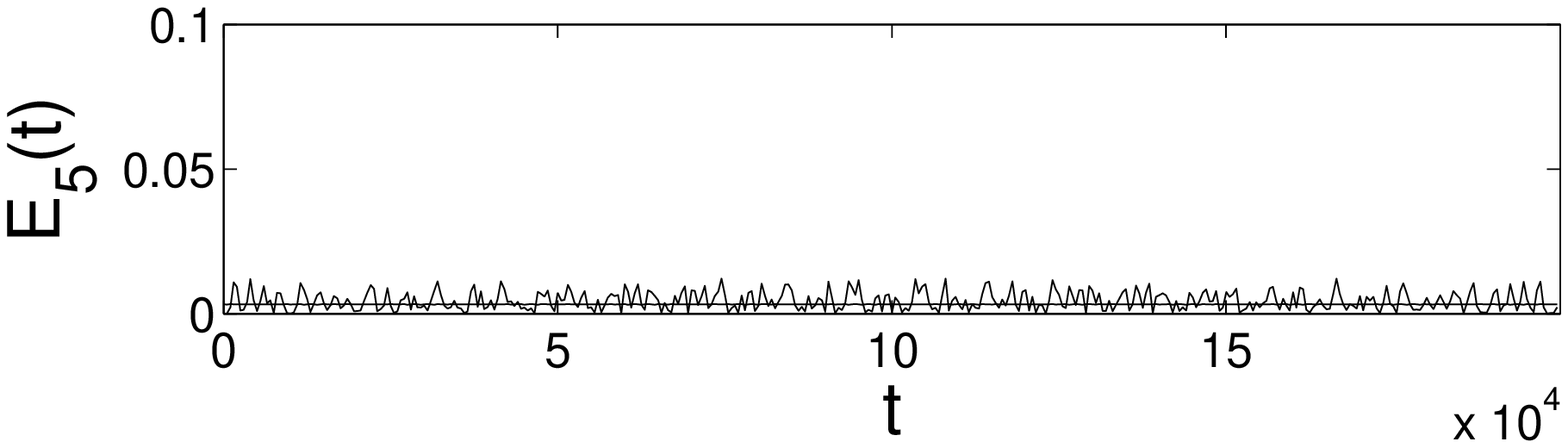}}}
{\caption{Evolution of the linear mode energies for the first five
modes on a large timescale for (i) the original FPU trajectory for
$\alpha=0.25, E=0.077, N=32$ \cite{fpu} (oscillating curves) and
(ii) the exact QB solution (almost straight lines).} \label{Fig1}}
\end{figure}

Let us consider the $\alpha$-model 
and start with showing the evolution of the original FPU trajectory
for $\alpha=0.25$, $N=32$ and an energy $E=0.077$ placed
initially into the mode with $q_0=1$. We plot the time dependence
of the mode energies $E_q(t)$ in Fig.\ref{Fig1} for the first
five modes. 
The period of the slowest ($q_0=1$) harmonic mode is 
$T_1 = 2\pi/\omega_1 \approx 66.02$.
We nicely observe  slow
processes of redistribution of mode energies, recurrences
and also even slower modulations of recurrence amplitudes
on time scales of the order of $10^5$.
Note also that on time scales comparable to $T_1$ all mode energies
show small additional oscillations - and it is easy to see that
they correspond to frequencies which are multiples of $\omega_1$.
The localization in $q$-space is also nicely observed, with 
the maximum of $E_5$ being eight times smaller than that of $E_1$.
  
In order to construct a $q$-breather, let us choose first $\alpha=0$,
excite a normal mode with $q=q_0$ to the energy $E_{q_0}=E$ and
let all other $q$-oscillators be at rest. With that we arrive at
a unique periodic orbit in the phase space of the FPU model. 
We expect that the orbit
will stay localized in $q$-space at least up to some critical
nonzero value of  $\alpha$ (and similarly for
the $\beta$-model \cite{symmanifolds}).
We proceed with a series of successful numerical experiments
continuing periodic orbits of the linear chain to nonzero
nonlinearity. These orbits as well as their Floquet spectra can be
calculated using well developed computational tools \cite{DBs} for
exploring periodic orbits.
\begin{figure}[p]
{\centering
\resizebox*{0.90\columnwidth}{!}{\includegraphics{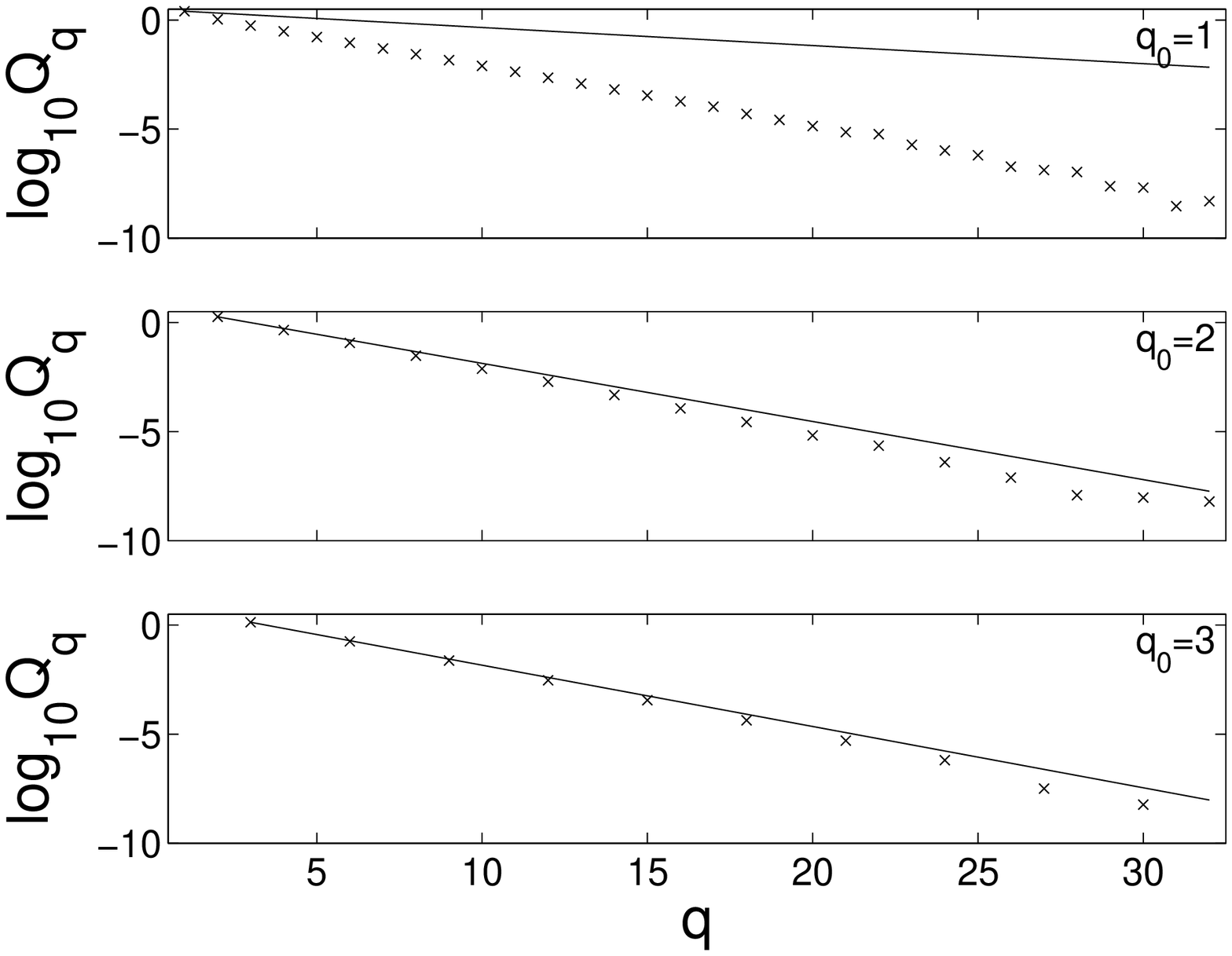}}}
{\caption{Snapshots of the linear mode coordinates $Q_q$
(symbols) along with analytical predictions (lines) for QBs with
$q_0=1,2,3$ for $\alpha=0.25, E=0.077, N=32$, at the moment when
all velocities $\Dot Q_q$ equal zero. Some modes have zero contributions
and their corresponding symbols are not plotted here.} 
\label{Fig2}} {\centering
\resizebox*{0.90\columnwidth}{!}{\includegraphics{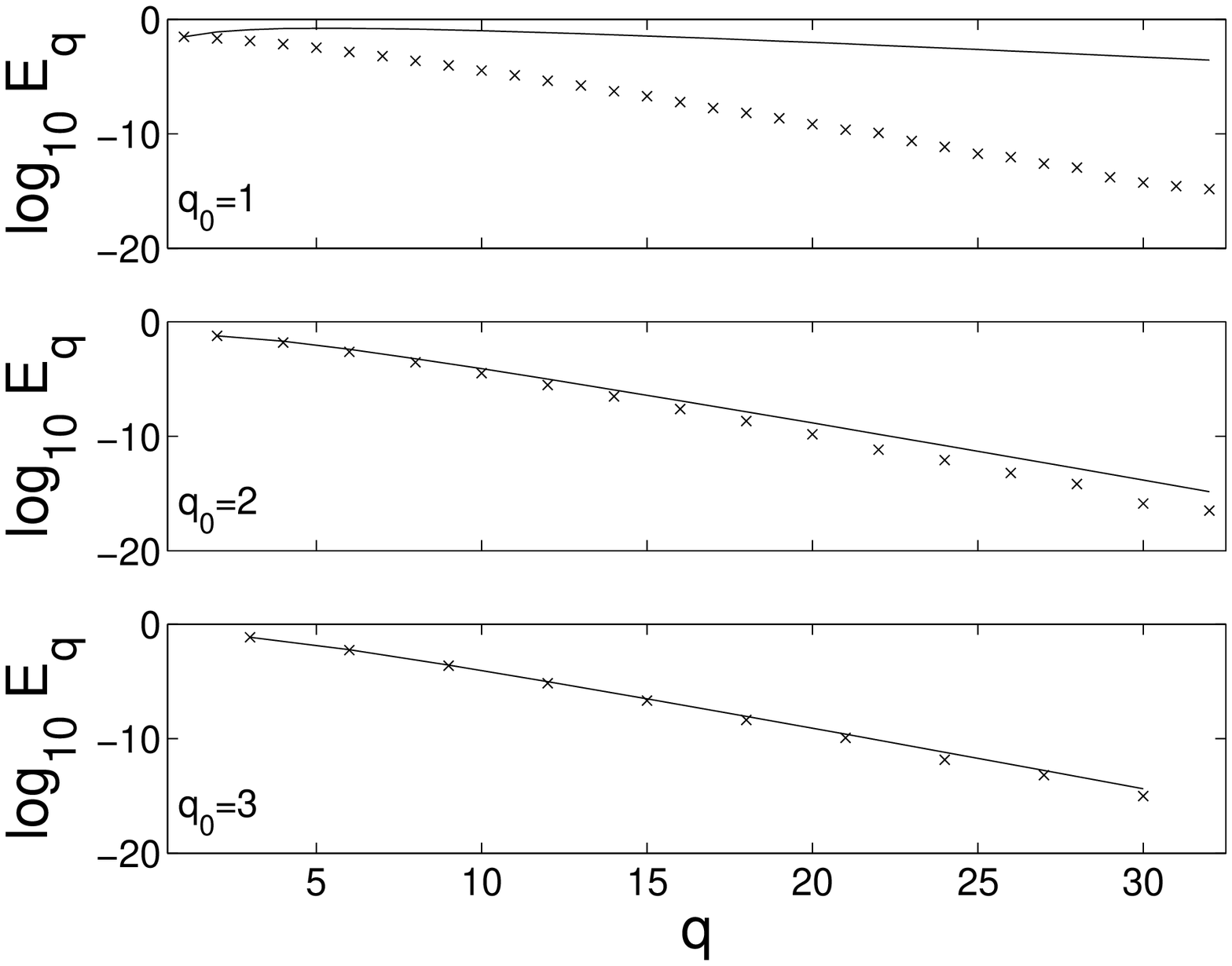}}}
{\caption{Distributions of the linear mode energies $E_q$ in
$q-$space for QBs with $q_0=1,2,3$ for $\alpha=0.25, E=0.077,
N=32$.}
\label{Fig3}}
\end{figure}
We choose a Poincar\'{e} section plane $\{x_s=0, \dot{x}_s>0\}$,
where $s=[2(N+1)/q_0]$ corresponds to an antinode of the mode
$Q_{q_0}$. We map the plane $\vec{y}$ (all phase variables
excluding $x_s$) onto itself integrating the equations of motion \eqref{eq1}
until the trajectory crosses the plane again:
$\vec{y}^{n+1}=\vec{\mathcal{F}}(\vec{y}^n)$. A periodic orbit of
the FPU chain corresponds to a fixed point of the generated map.
As the initial guess we use the point corresponding to the
$q_0$-th linear mode: $\Dot x_n(0)=\sqrt{\frac{2}{N+1}}\Dot
Q_{q_0}(0)\sin{\left(\frac{\pi q_0 n}{N+1}\right)}$, $x_n(0)=0$.
The vector function $\vec{\mathcal{G}}=\vec{\mathcal{F}}(\vec{y})
-\vec{y}$ is used to calculate the Newton matrix
$\mathcal{N}=\partial{\mathcal{G}(\vec{y})_i}/\partial y_j$. We
use a Gauss method to solve the equations
$\vec{\mathcal{G}}(\vec{y})={\mathcal{N}}(\vec{y} - \vec{y}\ ')$
for the new iteration $\vec{y}\ '$ and do final corrections to
adjust the correct total energy $E$. The iteration procedure continues
until the required accuracy $\varepsilon$ is obtained:
$||\vec{\mathcal{F}}(\vec{y})-\vec{y}||/||\vec{y}||<\varepsilon$
(we have varied $\varepsilon$ from $10^{-5}$ to $10^{-8}$), where
$||\vec{y}||=\max\{|y_i|\}$.
We also note that all periodic orbits computed here are invariant
under time reversal, which means that there exist times $t_0$
when $Q_q(t_0+t)=Q_q(t_0-t)$ for all $q$. Since mode velocities
vanish at these times, a QB can be also computed e.g. by initially
choosing all mode velocities to zero, and integrating until $\dot{Q}_{q_0}$
vanishes again at the same sign of $Q_{q_0}$ as it was at the starting point.
Then the fixed point of the corresponding map of the mode coordinates
only suffices to obtain an exact periodic solution.

\begin{figure}[p]
{\centering
\resizebox*{0.90\columnwidth}{!}{\includegraphics{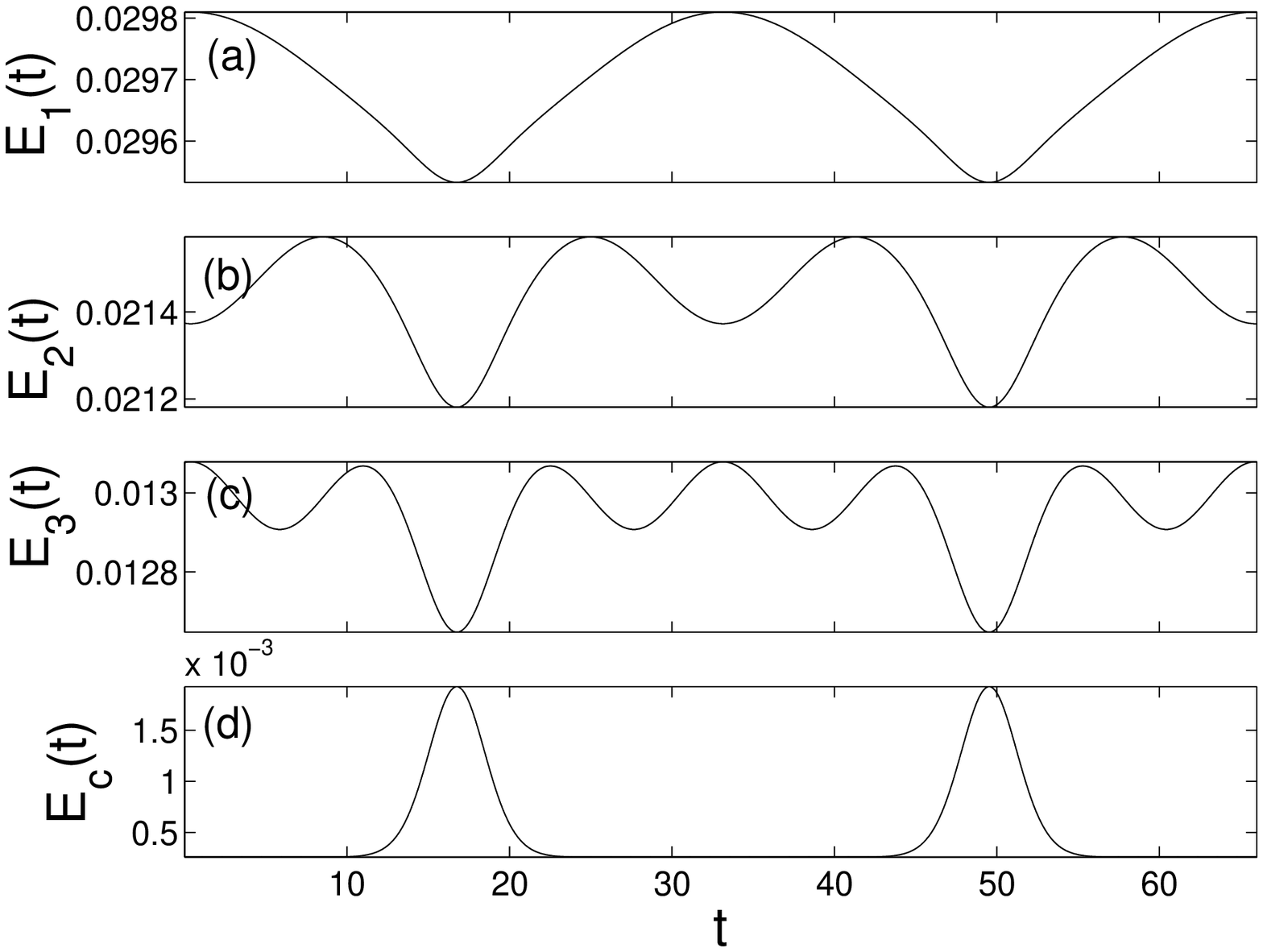}}}
{\caption{Evolution of the linear mode energies (a) $E_1$, (b)
$E_2$, (c) $E_3$ and the energy of nonlinear coupling (d) $E_c$
(see the text for definition) for the QB with $q_0=1, \alpha=0.25,
E=0.077, N=32$.} \label{Fig4}} {\centering
\resizebox*{0.90\columnwidth}{!}{\includegraphics{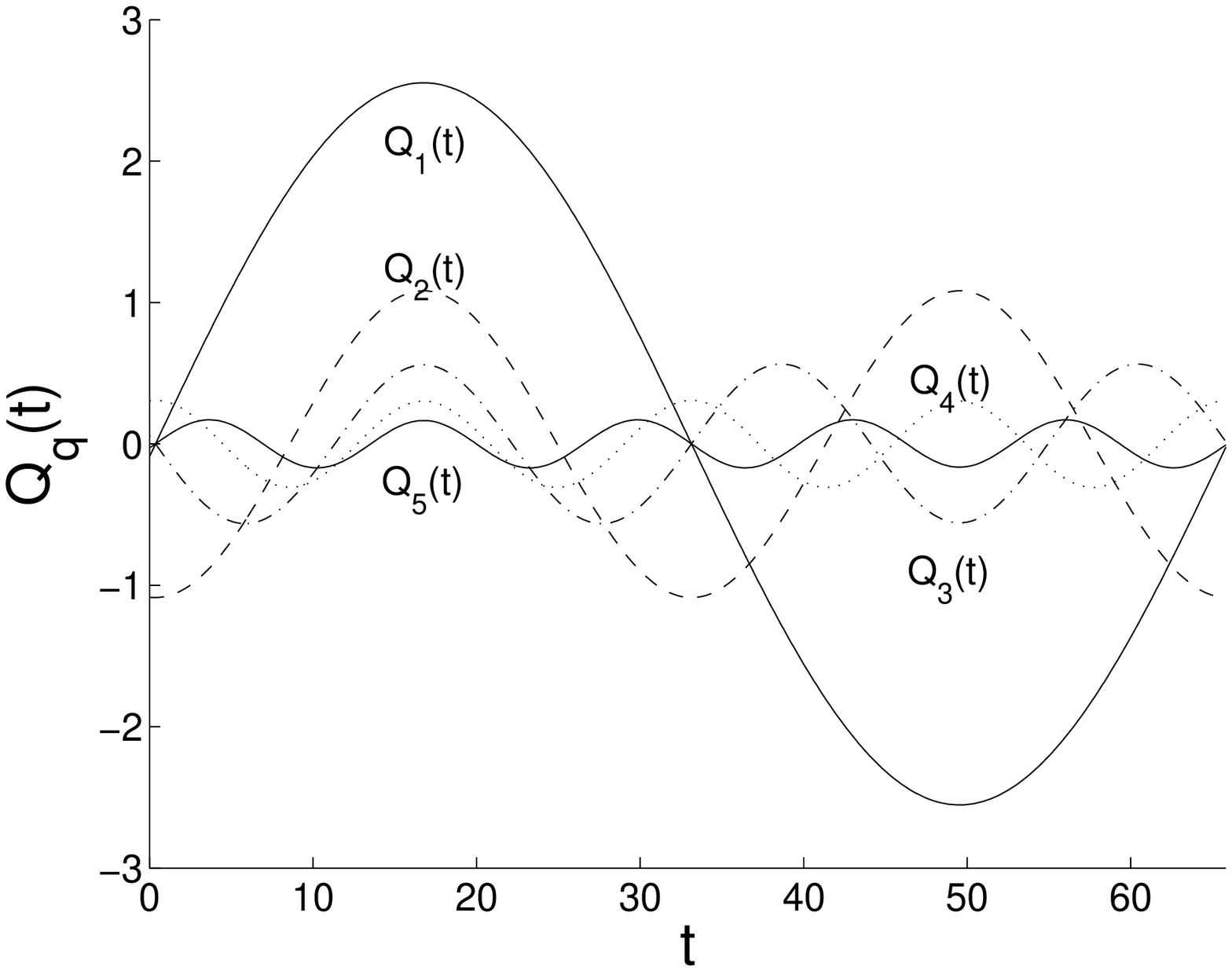}}}
{\caption{Evolution of the linear mode coordinates $Q_{1,2,3,4,5}$
 for the QB with $q_0=1, \alpha=0.25, E=0.077, N=32$.} \label{Fig5}}
\end{figure}

We have used one of the original parameter sets of the
FPU$-\alpha$ study $\alpha=0.25, E=0.077, N=32$ \cite{fpu} to find
stable $q$-localized QB solutions with most of the energy
concentrated in the mode $q_0=1$ (and added the cases $q_0=2,3$ for
comparison, see Fig.\ref{Fig2},\ref{Fig3}).
Note, that in \eqref{Bqlm} $2(N+1)$ is always a multiple of 2, and in
the particular case of $N=32$ it is also a multiple of 3. Then, for
$q_0=2$ and $q_0=3$ only modes with $q$ being a multiple of $q_0$ are
excited.
Let us discuss the properties of the found solutions in some detail. In
contrast to the original FPU trajectories, 
the QB is characterized by 
mode energies being almost constant in time. 
The straight lines in Fig.\ref{Fig1} compare the
mode energies on the QB with $q_0=1$ with the FPU trajectory.
The period of the QB solutions is very close to the corresponding period
$T_{q_0}=2\pi/\omega_{q_0}$ of the harmonic mode which is continued. 
During one QB period of oscillations a
relatively small energy interchange between the modes (of the order
of 2 percent) is observed 
(Fig.\ref{Fig4}(a-c)). 
with frequencies which correspond to multiples of the QB frequency -
just like the small fluctuations of the mode energies for
the FPU trajectory mentioned above.
There are well defined
intervals during which the energy of nonlinear coupling energy
$E_c=E_{tot}-\sum_{q=1}^{N}E_q$ ($E_{tot}$ being the full energy of the chain)
increases sharply, yet being overall very small (between
1 and 3 percent of the total energy, see Fig.\ref{Fig4}(d)).
In Fig.\ref{Fig5} we show the time dependence of the first five
mode coordinates on the QB with $q_0=1$. The time reversal symmetry
is nicely observed (although not used for construction) at
times $t_0 \approx 17,49$. Note also that while $Q_1$ oscillates
predominantly with the main QB frequency $\Omega_{QB}$ which
is close to $\omega_1$, the second mode $Q_2$ is dominated by
$2\Omega_{QB}$, the third one $Q_3$ by $3 \Omega_{QB}$ etc.

\begin{figure}[t]
{\centering
\resizebox*{0.90\textwidth}{!}{\includegraphics{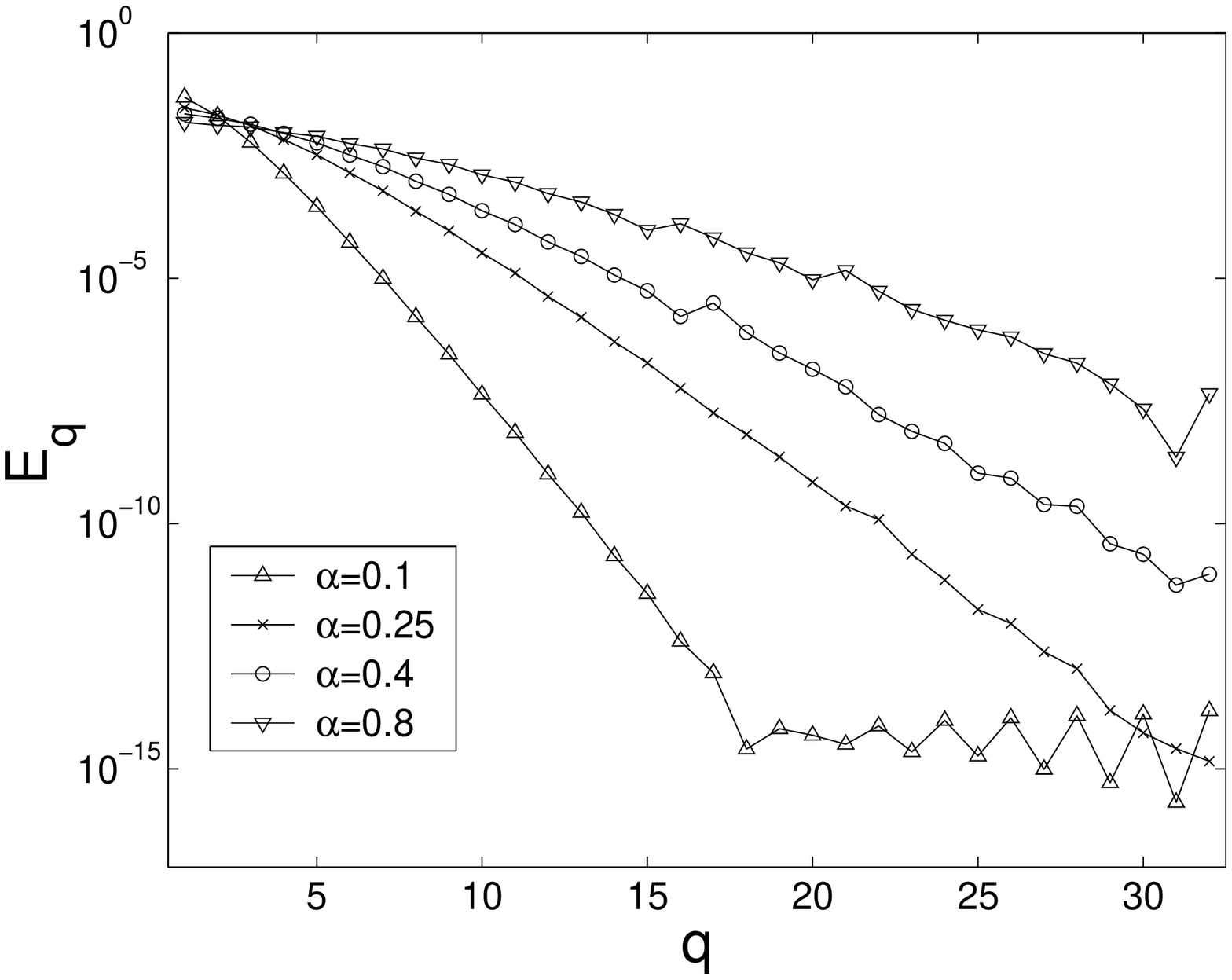}}}
{\caption{Distributions of the linear mode energies $E_q$ for $\alpha=0.1,
0.25, 0.4$ and $0.8$ for QBs with $q_0=1, E=0.077, N=32$. }\label{Fig6}}
\end{figure}

We conclude the numerical results on the $\alpha$-FPU case
with QB solutions for $q_0=1$, $E=0.077$ and various values of
$\alpha$ up to $\alpha=0.8$, which are shown in Fig.\ref{Fig6}.
We observe that the localization length increases with increasing
$\alpha$, so there is a clear tendency towards an equipartition
threshold.

\subsection{Estimating the localization length}

To demonstrate localization of a QB solution in $q$-space
analytically, we expand the solution to \eqref{alpmod} into an asymptotic
series with respect to the small parameter $\sigma=\alpha/\sqrt{2(N+1)}$:
\beq[eq_expn]
Q_q(t)=\sum_{n=0}^{\infty} \sigma^n Q^{(n)}_q(t)\mbox{.}
\eeq
Inserting this expansion into \eqref{alpmod}, we obtain equations
for the variables $Q^{(n)}_q(t)$. The equation for the zero order
$n=0$ reads
\beq[order0]
\Ddot Q^{(0)}_q+\om_q^2 Q^{(0)}_q=0
\mbox{,}
\eeq
and for $n>0$
\beq[ordern]
\Ddot Q^{(n)}_q+\om_q^2
Q^{(n)}_q=-\om_q \sum_{l,m=1}^N \om_l \om_m B_{qlm} \mspace{-18mu}
\sum_{\substack{n_{1,2}=0\\n_1+n_2=n-1}}^{n-1} \mspace{-18mu}
Q^{(n_1)}_l Q^{(n_2)}_m
\mbox{.}
\eeq

As the zero-order approximation we take a single-mode solution to \eqref{order0}
with $Q^{(0)}_q(t)\neq 0$ only for $q=q_0$:
\beq[sol0]
Q_{q}^{(0)}=\delta_{q,q_0} A_{q_0} \cos{\omega_{q_0} t}
\mbox{.}
\eeq

Consider the first-order equations (case $n=1$ in \eqref{ordern}). The
right-hand part in \eqref{ordern} contains only one nonzero term corresponding
to $l=m=q_0$, $n_1=n_2=0$. The coefficient $B_{qlm}$ here is non-zero only for
$q=q_0$ and $q=2q_0$ (here we assume $2q_0\le N$, so that the second Kronecker
symbol in \eqref{Bqlm} always equals zero). Thus, in the first order the
only variables different from zero are $Q^{(1)}_{q_0}(t)$ and
$Q^{(1)}_{2q_0}(t)$.

Similarly, for $n=2$, provided $3q_0\le N$, we get $Q^{(2)}_q(t)\neq 0$
for $q=q_0, 2q_0, 3q_0$ only.

The above allows us to formulate the following proposition.

In the $(k-1)$th order of asymptotic expansion, provided $kq_0\le N$, variables
$Q^{(k-1)}_q(t)$ differ from zero only for $q=q_0, 2q_0, \dots, kq_0$:
\begin{subequations}\label{Qzero}
\beq[Qzero_a]
Q^{(k-1)}_q(t)=0 \quad \forall q\notin \{q_0, 2q_0, \dots, kq_0\}
\mbox{.}
\eeq
It means, that the first non-zero expansion term for a mode $q=kq_0$ is
of the order $k-1$:
\beq[Qzero_b]
Q^{(m)}_{kq_0}(t)=0 \quad \forall m<k-1
\mbox{.}
\eeq
\end{subequations}

In Appendix \ref{alploc} we prove this statement by the method of mathematical
induction and approximate the first non-zero term for a mode $q=kq_0$ at
$kq_0\ll N$:
\begin{subequations}\label{senior}
\beq[senior1]
Q^{(k-1)}_{kq_0}(t)=A_{kq_0}\left(\cos k\om_{q_0}t
+O\left((kq_0/N)^2\right)\right)
\mbox{,}
\eeq
where
\beq[senior2]
A_{kq_0}=\frac{A_{q_0}^k}{\om_{q_0}^{k-1}}
\mbox{.}
\eeq
\end{subequations}

Ignored expansion terms lead to shifting the QB orbit frequency and
next-order corrections to its shape.

Multiplying \eqref{senior} by $\sigma^{k-1}$ and inserting it into \eqref{Eq},
we approximate the mode energies as
\begin{subequations}\label{Ekq0}
\beq[Ekq0a]
E_{kq_0}=k^2\gamma^{k-1} E_{q_0} \cdot
\left(1+O\left((kq_0/N)^2\right)\right)
\left(1+O(\sigma)\right)
\mbox{,}
\eeq
where
\beq[Ekq0b]
\gamma=\frac{\alpha^2 (N+1)^3 E_{q_0}}{\pi^4 q_0^4}
\mbox{.}
\eeq
\end{subequations}
Note, that although mode energies are not strictly conserved in time, their
variation is small, being limited to the orders of magnitude indicated in
\eqref{Ekq0a} in parentheses (cf.~Fig.\ref{Fig4}).
We arrive at an exponential decay in $q$-space dressed with a
power law:
\begin{equation}
\ln(E_q) = 2 \ln (\frac{q}{q_0}) + \frac{q}{q_0} 
\ln \gamma +\ln \left( \frac{E_{q_0}}{\gamma}\right) 
\label{alpha-logarithm}
\end{equation}
with exponent $q_0^{-1}\ln \gamma$ for the exponential part.
The predicted decay (\ref{Ekq0}) fits nicely with
a computed QB with $q_0=1$ and $\alpha=0.025$ as shown in \cite{sfmvioik05}.
In Figs.\ref{Fig2},\ref{Fig3} we compare (\ref{Ekq0}) with the numerical
results for $\alpha=0.25$ and $q_0=1,2,3$. While $q_1$ shows that
the analytical results underestimate the degree of localization
with increasing $\alpha$, we note that it does even at these values
$\alpha$ better for larger values of $q_0$. 

The calculation described is expected to fail at $\gamma$ close
to or greater than one, when \eqref{Ekq0} does not support
$q$-space localization. At the same time the comparison
with the numerical results shows that higher order corrections
to the analytical decay law extend the region of QB localization.
Thus we estimate a lower bound for the QB delocalization threshold 
energy $E^{loc}$ as
\beq[alp_deloc]
 E^{loc} =\pi^4 \alpha^{-2} (N+1)^{-3}
q_0^4 \mbox{.}
\eeq
The scaling behaviour is in good agreement with an analytical
estimation of the equipartition threshold in the
$\alpha$-model due to second order nonlinear resonance overlap, suggested in
\cite{Shepel}, which reads $\alpha^2 E N^3 \approx q_0^4$. 

Finally we stress an even closer relation between the localization of $q$-breathers
within the used perturbation theory
and the Fourier series convergence of analytic periodic functions 
(\ref{aho}),(\ref{ahosolution}),(\ref{ahofc}). One arrives from the QB problem
to these equations by simply assuming $\omega_q=const$.  

\subsection{Stability of $q$-breathers in the $\alpha$-FPU system}

An important problem is the stability of QBs. For computing linear stability of an
orbit $\hat Q_q(t)$, the phase space flow around it is linearized by making a
replacement
\beq[linrepl]
Q_q=\hat Q_q(t)+\xi_q
\eeq
in the equations of motion \eqref{betmod} and subsequent linearizing the
resulting equations with respect to $\xi_q$. Orbit stability is
then characterized by the eigenvalues of the Floquet matrix, which defines the
linear transformation of small deviations $\xi_q$ by the linearized equations
over one period of the orbit. If all eigenvalues $\mu_j$ have the absolute value
one, the orbit is stable, otherwise it is unstable \cite{DBs,DBs2}.

All QB solutions presented in this section are linearly stable,
i.e. all eigenvalues of the Floquet matrix which characterizes
the linearized phase space flow around the QB orbits reside
on the unit circle. This is at variance with
the  case of the $\beta$-FPU model, which will be discussed below.

\section{$\beta$-FPU system}

\subsection{Numerical results}

Launching an FPU trajectory by exciting a single low-frequency mode
leads to similar observations as for the $\alpha$-model. Again energy is
localized in $q$-space on a few modes, sometimes coined natural packets \cite{lbagsp04},
which again persists for very long times.

It is possible to construct QBs in the $\beta$-FPU model,
following the same way as described above for the
$\alpha$-model.

For numerical calculation we use a modified scheme. The section
plane is defined in the $q$-space as $\{Q_{q_0}=0, \Dot
Q_{q_0}>0\}$. It is parametrized as $\vec{r}\equiv \{ \dot{Q}_q,
q\neq q_0\}$. The even coupling potential and fixed boundary
conditions enable us to introduce an additional constraint
$Q_q(t=0)=0$, and the velocity $\Dot{Q}_{q_0}$ is obtained using the
condition of energy conservation $\dot{Q}_{q_0}(t=0)=
\sqrt{2E-\sum_{q \neq q_0}\dot{Q}_q^2(t=0)}$. As in the case of
the $\alpha$-model, the QB is searched as a fixpoint of the
mapping $\vec{r}^{n+1}=\vec{\mathcal{F}}(\vec{r}^n)$.

\begin{figure}[t]
{\centering
\resizebox*{0.90\columnwidth}{!}{\includegraphics{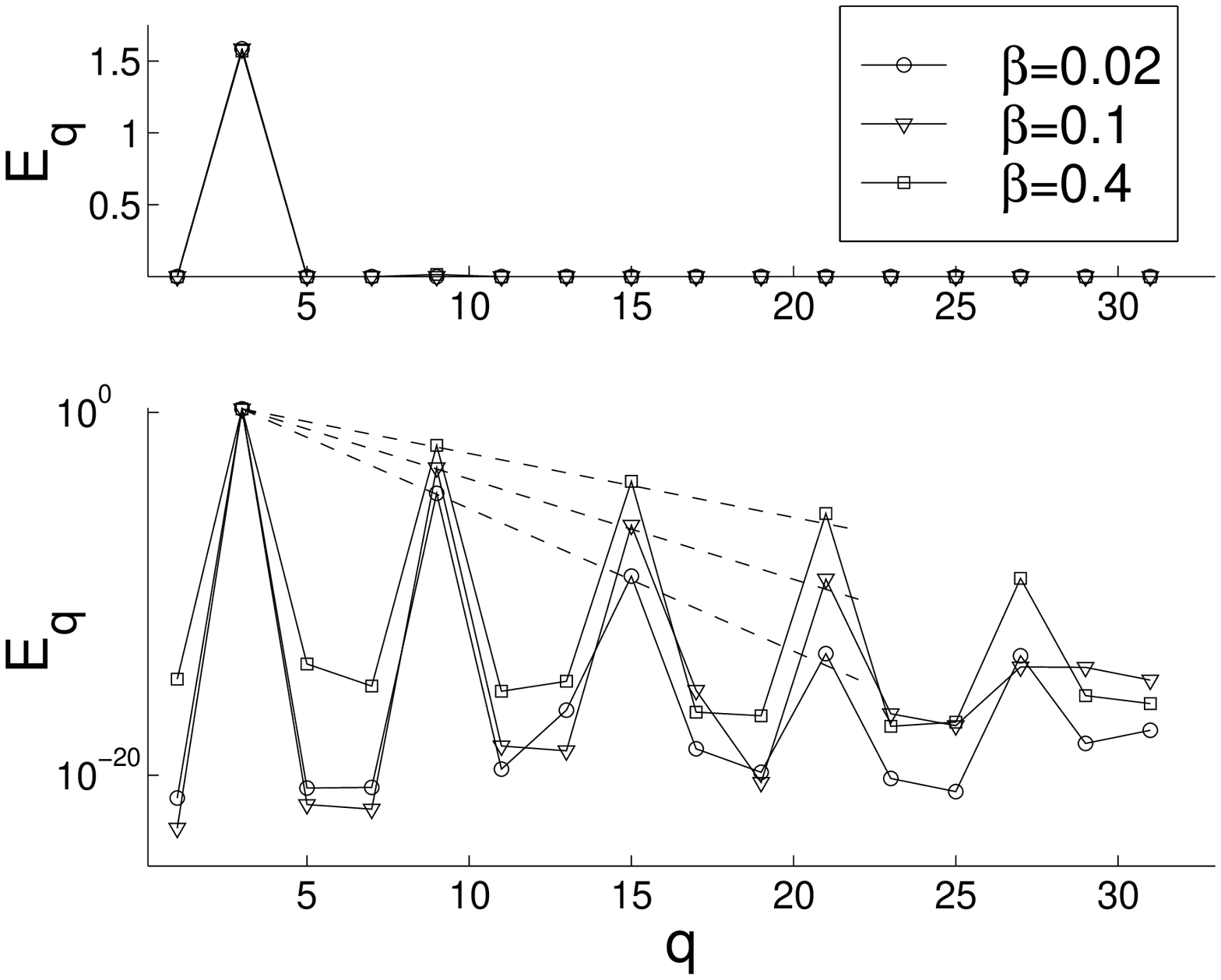}}}
{\caption{Energy distributions between $q$-modes in QBs for
different nonlinear coupling coefficients $\beta$ versus $q$ in
linear and log scales with analytical estimations of the QBs
exponential localization (dashed lines). Parameters are
$E=1.58$, $q_0=3$, $N=32$. Only odd modes are shown (see text).
The symbols for $q \neq 3,9,15,21,27$ represent upper bounds,
the real mode energies might be even less. Note that QBs persist
even far beyond the stability threshold (see Fig.\ref{oldfig2}).
}\label{oldfig1}}
\end{figure}

We obtain QBs which are exponentially localized in $q$-space
(Fig.\ref{oldfig1}). The smaller $\beta$, the faster is the
decay of the energy distribution with increasing wave number
$q$. Note, that due to the parity symmetry of the $\beta$-model
(Eq.(\ref{eq2}) is invariant under $x_n \rightarrow -x_n$ for
all $n$) only odd $q$-modes are excited by the $q_0=3$ mode and
get coupled \cite{bushes}. This follows also from the coupling matrix
(\ref{Cqlmn}).

\subsection{Estimating the localization length}

In the analytical computation the solution to \eqref{betmod} is
expanded in powers of a small parameter
$\rho=\frac{\beta}{2(N+1)}$. In the $n$-th order of expansion
variables $Q^{(n)}_q$ differ from zero at $q=q_0, 3q_0, \dots,
(2n+1)q_0$ only. Then the first non-zero expansion term for a
mode $(2n+1)q_0$ is of the order $n$. Using the same approach of
mathematical induction as described in Appendix \ref{alploc},
mode energies in a QB are approximated as follows:
\begin{subequations}\label{bE}
\beq[bEa]
E_{(2k+1)q_0}=\lambda^k E_{q_0} \cdot
\left(1+O\left(((2k+1)q_0/N)^2\right)\right)
\left(1+O(\rho)\right)
\mbox{,}
\eeq
where
\beq[bEb]
\lambda=\frac{9\beta^2 E_{q_0}^2 (N+1)^2}{64\pi^4 q_0^4}
\mbox{.}
\eeq
\end{subequations}
Again, time variation of mode energies is limited to the orders of magnitude
indicated in \eqref{bEa}. 
We arrive at a pure exponential decay
\begin{equation}
\ln (E_q) =  \ln E_{q_0} +\frac{1}{2}(\frac{q}{q_0}-1)\ln \lambda
\end{equation}
with exponent $(2q_0)^{-1} \ln \lambda $.
The predicted decay (\ref{bE}) fits nicely with the computed QBs
in Fig.\ref{oldfig1}.

The QB delocalization threshold is then estimated as
\beq[bet_deloc]
E_{q_0} =\frac{8}{3}\pi^2 q_0^2 \beta^{-1} (N+1)^{-1}
\mbox{.}
\eeq
The scaling behaviour is in good agreement with an analytical estimation
of the equipartition threshold in the $\beta$-model due to second order
nonlinear resonance
overlap, obtained in \cite{Shepel}, which reads
$\beta E N \approx q_0^2$ .

\subsection{Stability of $q$-breathers for the $\beta$-FPU model}

\begin{figure}[t]
{\centering
\resizebox*{0.90\columnwidth}{!}{\includegraphics{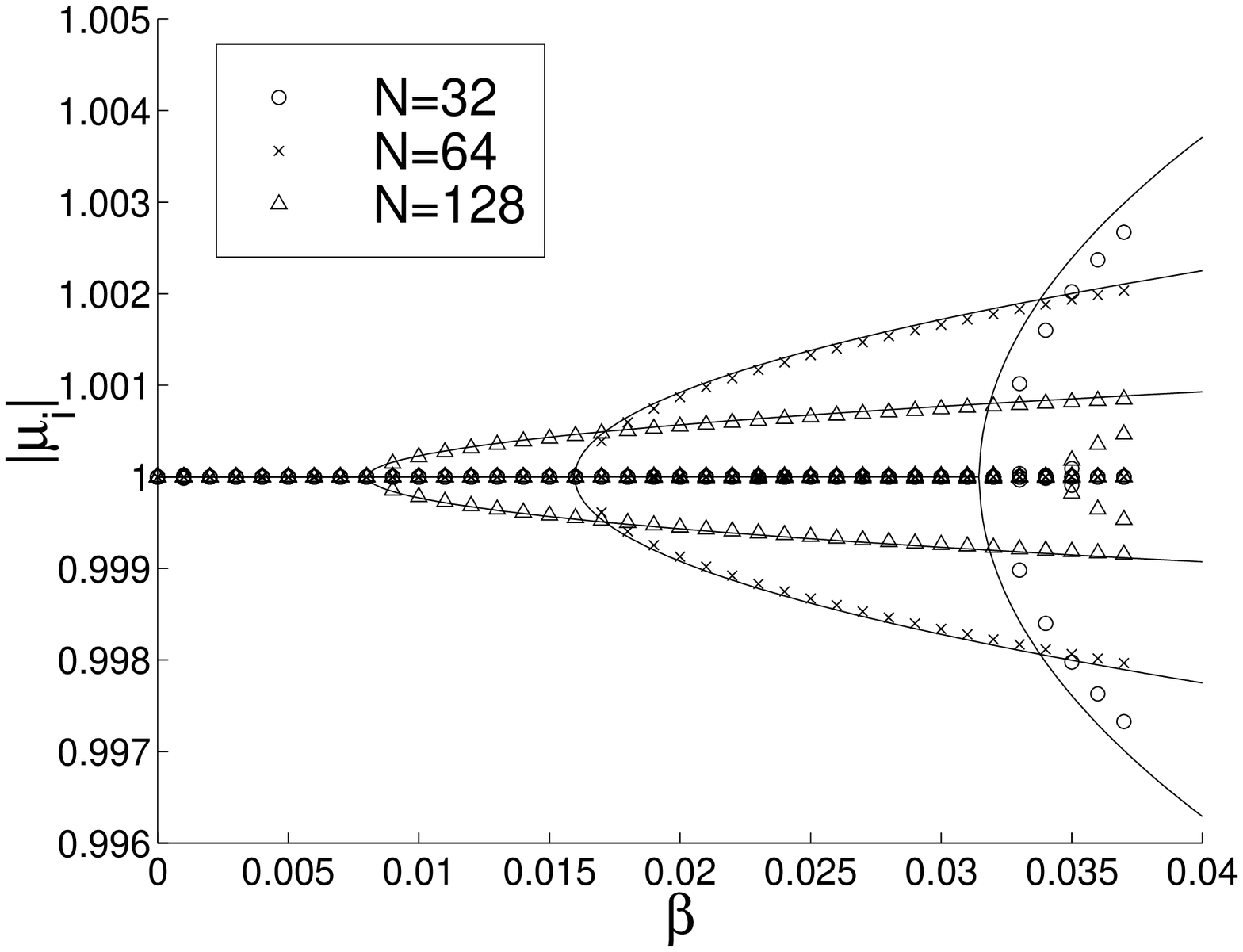}}}
{\caption{Absolute values of Floquet multipliers $|\mu_i|$ of
QBs with the energy $E=1.58$ and $q_0=3$ and different $N$
versus $\beta$. Symbols: numerical results, lines: analytical
results.} \label{oldfig2}}
\end{figure}

First, we study QB stability by computing the Floquet matrix
numerically and diagonalizing it. In Fig.\ref{oldfig2} we plot
the absolute values of the Floquet eigenvalues of the computed
QBs versus $\beta$ for different system sizes $N$. QBs are
stable for sufficiently weak nonlinearities (all eigenvalues
have absolute value 1). When $\beta$ exceeds a certain threshold
two eigenvalues get absolute values larger than unity (and,
correspondingly, another two get absolute values less than
unity) and a QB becomes unstable. Remarkably, unstable QBs can
be traced far beyond the stability threshold, and moreover, they
retain their exponential localization in $q$-space
(Fig.\ref{oldfig1}). As $\beta$ is increased further, new
bifurcations of the same type are observed.

To study QB stability analytically in the first-order
approximation, we write down the QB solution in the form
\beq[QB] \hat Q_q(t) = \delta_{q q_0}
A\cos\hat{\omega}t+O(\rho)\mbox{,} \eeq where $\hat{\omega}$ is
the QB frequency, slightly shifted from $\om_{q_0}$ due to
nonlinearity. The residual term $O(\rho)$ includes corrections
to the QB orbit shape by expansion terms of order one and
higher.

The first-order correction to the QB frequency is determined by secularity
caused by resonant nonlinear self-forcing of the mode $q_0$ in the first order
of expansion:
\beq[eq_q0]
\Ddot Q^{(1)}_{q_0}+\om_{q_0}^2 Q^{(1)}_{q_0}=-3 \om_{q_0}^4 {Q^{(0)}_{q_0}}^3
\eeq
This equation is identical to that of an isolated oscillator with cubic
nonlinearity (Duffing oscillator). The well-known expression of nonlinear
frequency shifting in the Duffing oscillator then yields
\beq[eq_Omg]
\hat{\omega} =
\omega_{q_0} \left( 1+\frac{9\beta E_{q_0}}{8(N+1)}+O(\rho^2) \right)
\eeq

Linearizing equations of motion \eqref{betmod} around \eqref{QB} according to
\eqref{linrepl}, we arrive at a Mathieu equation (see Appendix \ref{betstab}),
which finally leads to an estimation of the Floquet multipliers which leave the
unit circle due to a primary parametric resonance and cause instability:
\beq[mults]
|\mu_{j_1 j_2}|= 1\pm\frac{\pi^3}{4(N+1)^2}\sqrt{R-1+O(\frac{1}{N^2})}
\eeq
where 
\begin{equation}
R=6\beta E_{q_0}(N+1)/\pi^2\;.
\label{R}
\end{equation}
The bifurcation occurs at
$R=1+O(1/N^2)$. This instability threshold coincides with the criterion of
transition to weak chaos reported by De Luca et al. \cite{deLuca}.
Note, that \eqref{mults} does not contain the principal mode number $q_0$.
Below the stability threshold (except for possible small high-order resonance
zones) a set of stable QB modes exists. In the thermodynamical limit
$N\to\infty$, however, the energy of stable QBs tends to zero.
Note that according to our analytical (Appendix \ref{betstab}) and numerical results
the instability modes correspond to $q'=q_0\pm 1$ and are even modes if $q_0$ is odd
and vice versa. The observed instability is thus connected to a lowering of the symmetry
of the higher symmetry QB. That has been also observed to be the driving pathway for
the onset of low-dimensional stochasticity in the FPU trajectory at the weak chaos transition
\cite{deLuca},\cite{jdlajlsr95}, where the FPU trajectory acquires chaotic components in the time evolution,
while still being localized in $q$-space.

The result \eqref{mults} is plotted in Fig.\ref{oldfig2} with
solid lines for $N=32$, 64 and 128, demonstrating good agreement
with the numerical results. The agreement improves with
increasing $N$ \cite{corrections}.

Driscoll and O'Neil studied the instability of a single soliton in the continuum
mKdV limit of the $\beta$-FPU model \cite{cfdtmon76} with periodic boundary conditions.
A stability threshold obtained within the mKdV equation
will qualitatively or semiquantitatively agree with the correct value obtained for the discrete chain,
if the instability sets in for QBs which do not contain significant short wavelength components.
The relation between single soliton and QB stability is less clear, since in the
limit of vanishing nonlinearity QBs in chains
or field equations with periodic boundary conditions correspond to standing waves, while
single solitons transform into plane (running) waves.

We conclude this section with the observation that the stability (\ref{R}) and delocalization 
(\ref{bet_deloc}) threshold
estimates for QBs contain a unique parameter $\beta E N$ where $E$ is the total energy,
at variance with scaling estimates for the transition times to equipartition \cite{jdlajlsr99},
which obtain $\beta E/N$ instead.

\section{QBs and the FPU-trajectory for the $\alpha$-FPU model}

\begin{figure}[t]
{\centering
\resizebox*{0.90\columnwidth}{!}{\includegraphics{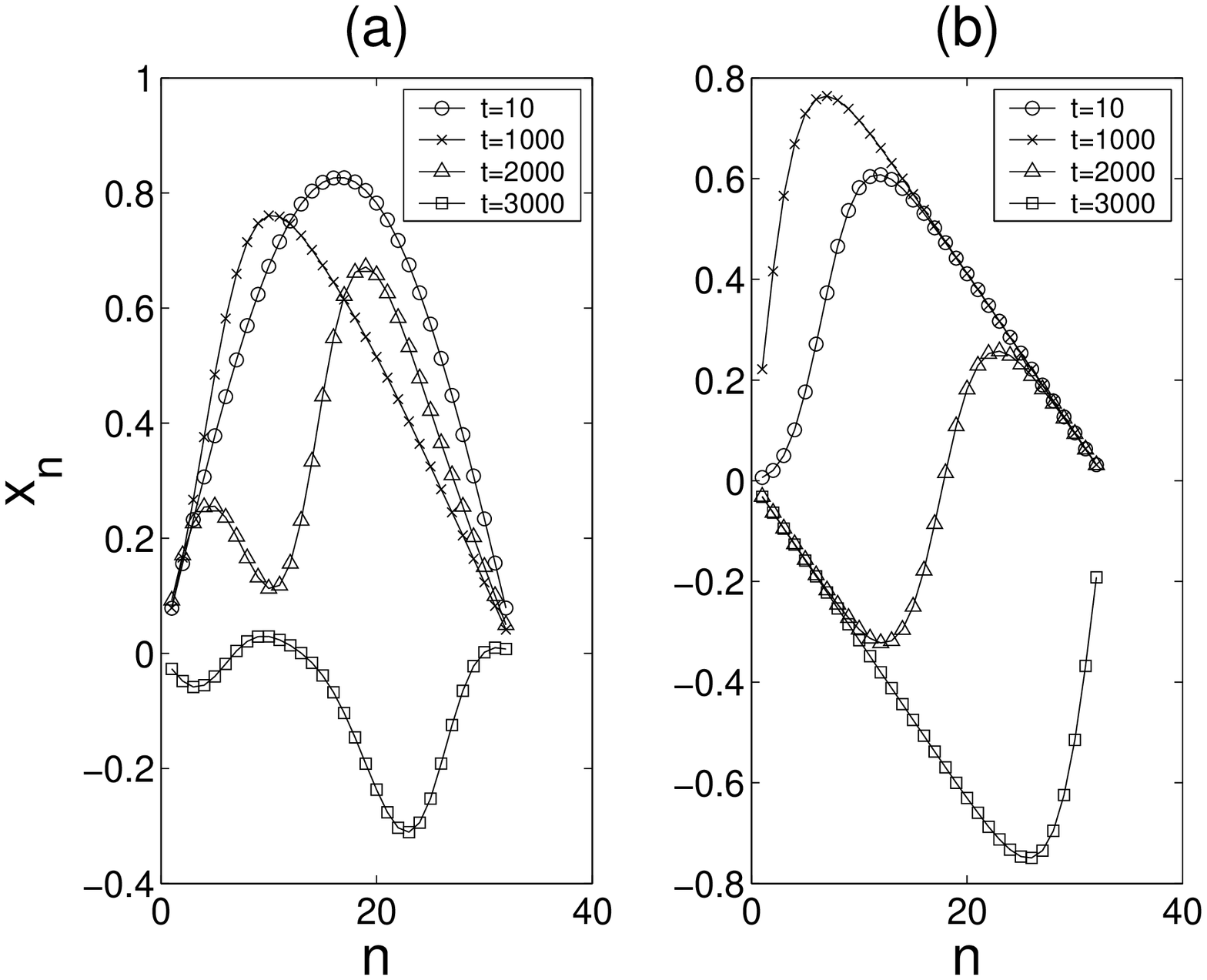}}}
{\caption{Snapshots of displacements (a) of the original FPU
trajectory for $\alpha=0.25, E=0.077, N=32$ \cite{fpu} and (b) of
the corresponding exact QB solution from Fig.\ref{Fig2} taken at
different times. } \label{oldfig4}}
\end{figure}

Departing from the the QB orbit in phase space in the direction
of the initial condition of the FPU trajectory implies adding
to the QB solution, which is localized in $q$-space, a perturbation,
which is localized in q-space as well. The perturbed trajectory
will evolve essentially on a low-dimensional torus in phase space,
whose dimension will correspond to the number of modes excited
on the QB orbit - e.g. for $\alpha=0.25$ and $q_0=1$ about four or five.
In Fig.\ref{oldfig4} we compare snapshots of displacements at
different times obtained for the original FPU trajectory in
\cite{fpu} and for the numerically exact QB solution from
Fig.\ref{Fig2} for $\alpha=0.25$ and observe
similar evolution patterns (see also Ref. \cite{lcmcsmpegdc97}). 
Moreover, we took a series of points
on a line which connected initial conditions of the FPU trajectory
($E_{q \neq 1} =0$) with the numerically exact QB solution from
Fig.\ref{Fig2}. For each of these points we integrated the
corresponding trajectory and measured  the average deviation
$\Delta$ from the QB orbit. The dependence of $\Delta$ on the line
parameter turns out to be an almost linear one, starting from zero
when being very close to the QB orbit, and ending with a maximum
value when being close to the FPU trajectory. That supports the
expectation that the FPU trajectory is a perturbation of the QB
orbit. The FPU recurrence is gradually appearing with increasing
$\Delta$ and is thus
directly related to the regular motion of a slightly perturbed QB
periodic orbit, which we tested also numerically.
\begin{figure}[t]
{\centering
\resizebox*{0.90\columnwidth}{!}{\includegraphics{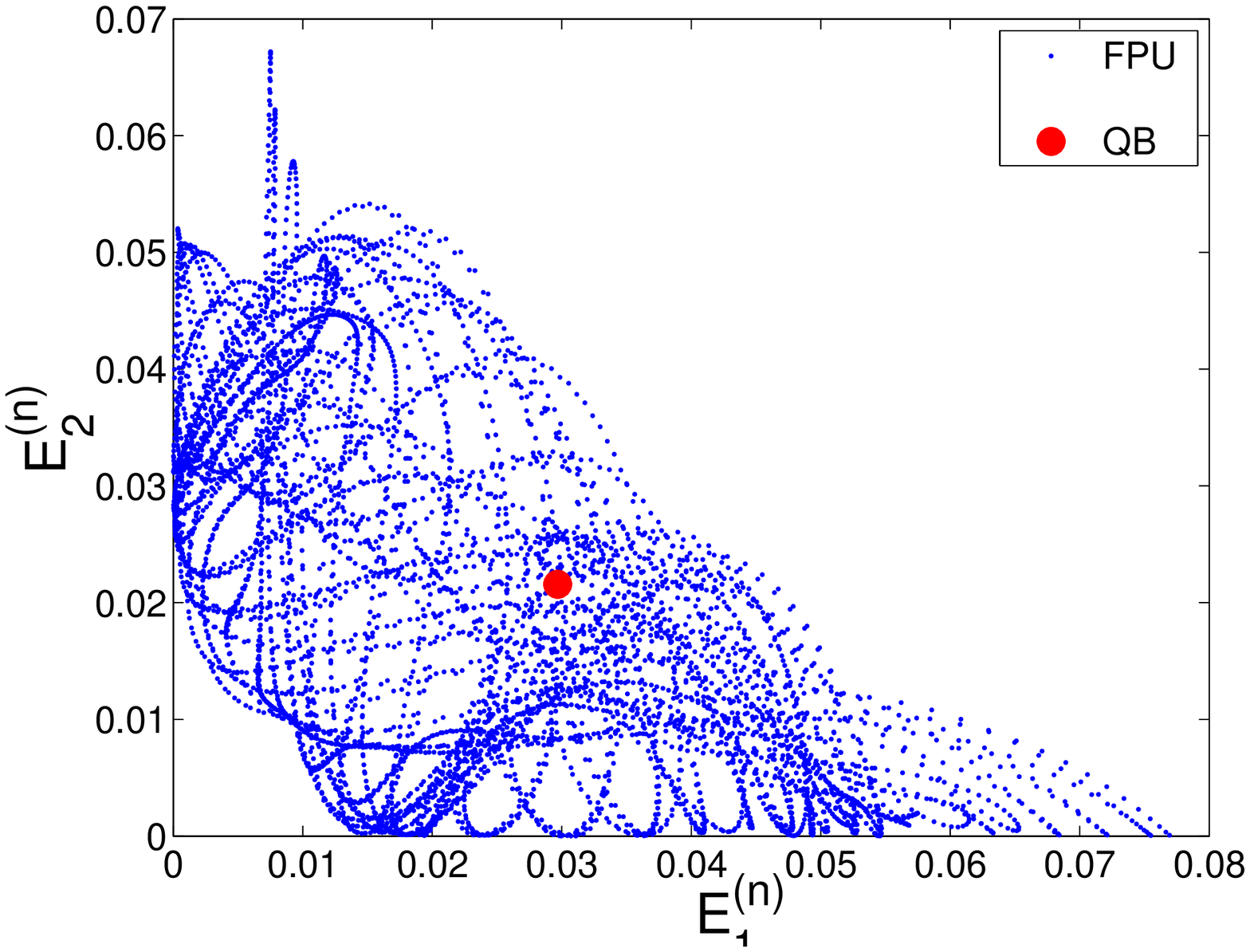}}}
{\caption{Map of the mode energies $E_1$ and $E_2$ 
on the FPU trajectory after
consecutive periods of the corresponding $q$-breather for
$N=32$, $\alpha=0.25$ and $q_0=1$. The thick dot is the
results for the QB orbit.} \label{poincarefig}}
\end{figure}
In Fig.\ref{poincarefig} we plot the mode energies $E_1$ and $E_2$
during integration of the FPU trajectory for $\alpha=0.25$ and 
$q_0=1$ after consecutive periods of the $q$-breather solution.
The regular pattern indicates regular motion, and the thick dot,
which corresponds to the $q$-breather solution itself, resides inside
the quasiperiodic cloud of the FPU trajectory, indicating once
more that the FPU trajectory is a perturbation of the $q$-breather
and evolves around the QB in phase space. 
Zooming the time dependence of the mode energies for the FPU trajectories
on time scales comparable to the QB period shows very similar nearly periodic
fluctuations as in Fig.\ref{Fig4} which are generated by the QB period.
Using the linearized phase space flow around a QB we can estimate an effective
recurrence time, which for the original FPU case is two times smaller than the recurrence
time for the FPU trajectory. We tracked the change of the recurrence time with increasing
$\Delta$. When coming closer to the FPU trajectory, simply every second recurrence as observed
for small deviations from the QB is suppressed, leaving us with the FPU recurrence time.
That effect may be due to additional nonlinear contributions to the phase space flow around 
a QB.

Finally we note that extremely
long computations of the FPU trajectory have been reported recently \cite{agsptp05}.
The trajectory localizes in $q$-space (and thus stays close to a $q$-breather)
for times up to $10^{10}$. Only after that a mixing of mode energies is observed,
possibly due to Arnold diffusion. Notably that critical time has been estimated by numerical
scaling analysis for shorter transition times \cite{lcmcsmpegdc97}.

\section{Towards transient processes and thermal equilibrium}

Once the existence and stability of QBs as exact solutions are
established, it is interesting to analyze the contribution of
these trajectories to the dynamics of transient processes
and thermal equilibrium. 
It is well-known, that in states, corresponding to energy
equipartition between degrees of freedom (or thermal equilibrium)
energy distribution demonstrates strong statistical properties,
with no energy concentrations on average in some subparts
of phase space. This circumstance,
however, does not forbid existence of finite-time energy
localizations whose lifetime may substantially exceed the
characteristic period of plane waves. We recall the 
studies of discrete breather contributions to
the dynamics of nonlinear lattices in thermal equilibrium
and transient processes \cite{statistics}.

Here we present results of numerical simulations of the
$\beta$-FPU chain ($N=100$, fixed boundary conditions) with two
types of initial conditions: (i) all energy is located in a single mode
$E_{tot}=E_{3}=1.58, E_{q\neq3}=0$, (ii) all energy is randomly
distributed among all modes: $Q_q(0)=\xi_k/\omega_q, \
\dot{Q}_q(0)=\eta_k$, where $\xi_k, \eta_k$ are random numbers,
uniformly distributed in $[-c,c]$, $c$ taken to ensure
$E_{tot}/N=0.2$.

\begin{figure}[p]
\includegraphics[width=0.47\textwidth]{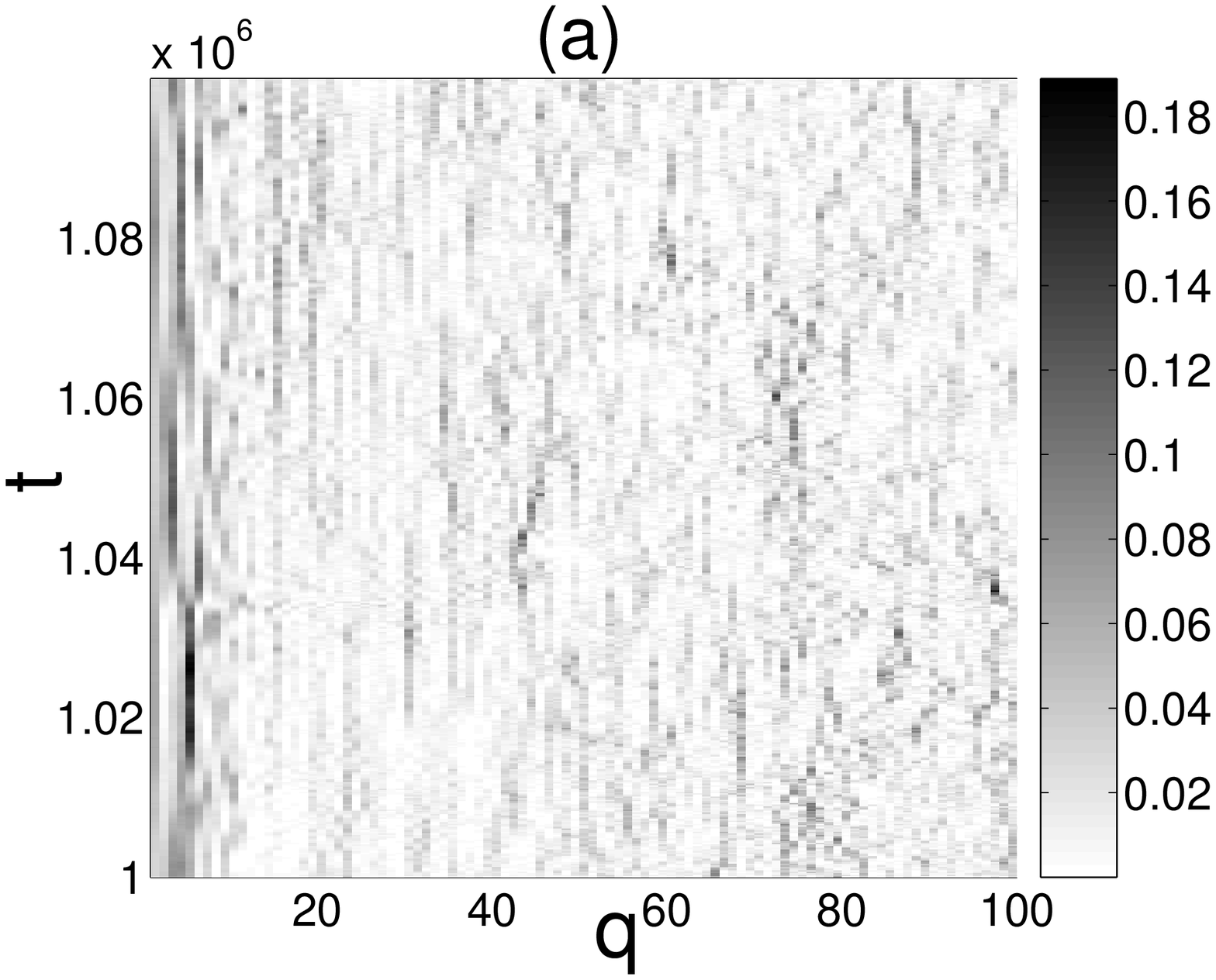}
\hfill
\includegraphics[width=0.47\textwidth]{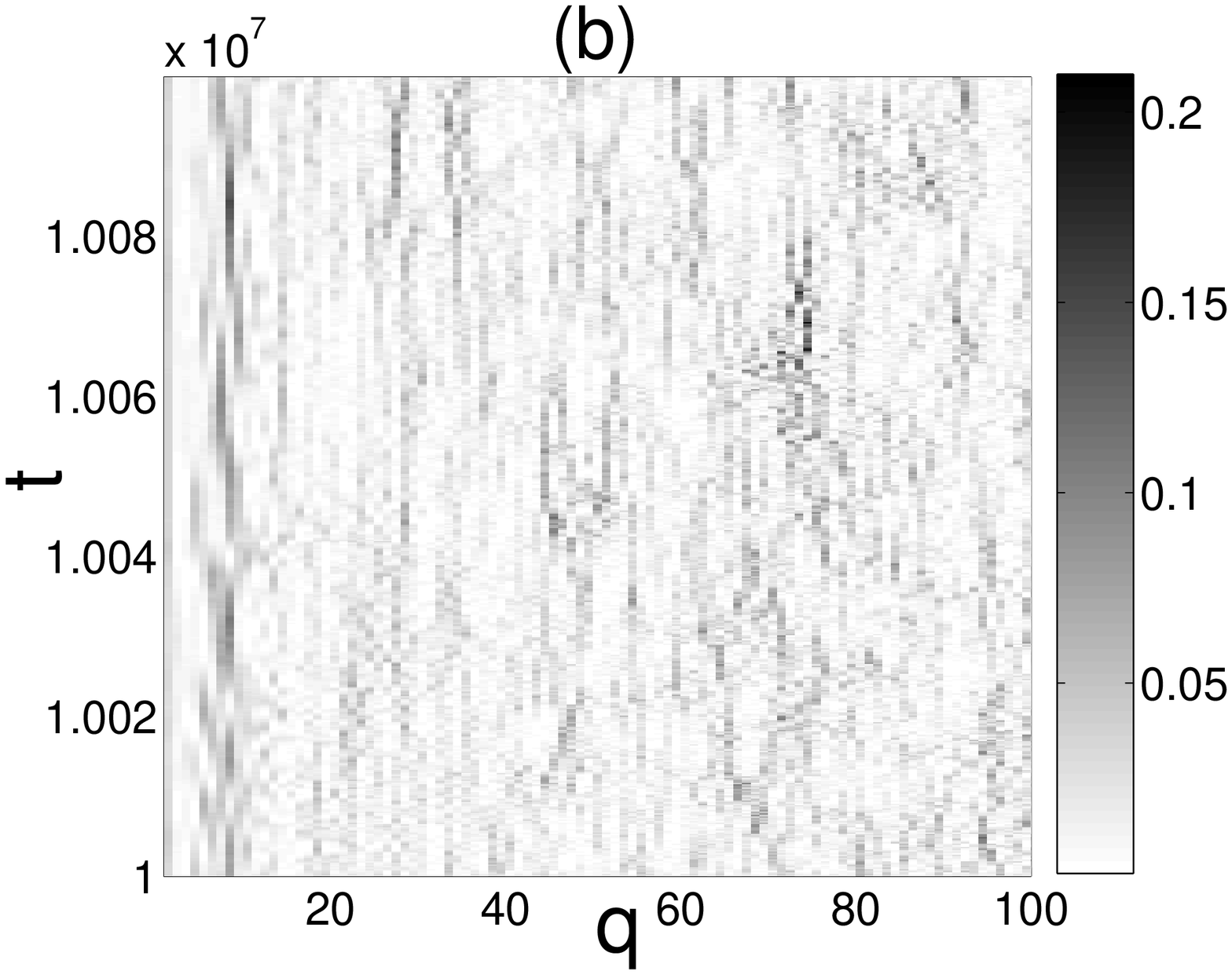}
\\
\includegraphics[width=0.47\textwidth]{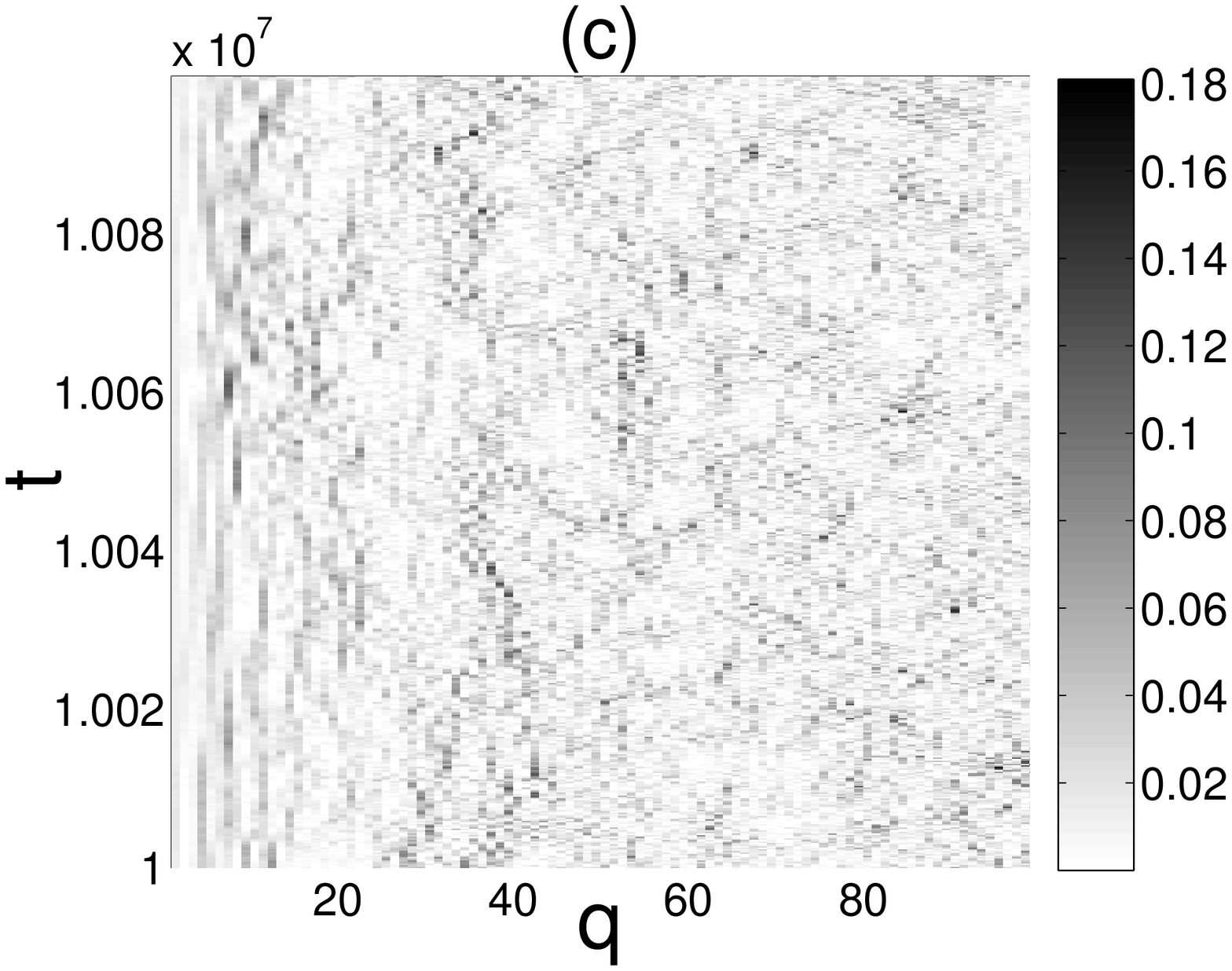}
\hfill
\includegraphics[width=0.47\textwidth]{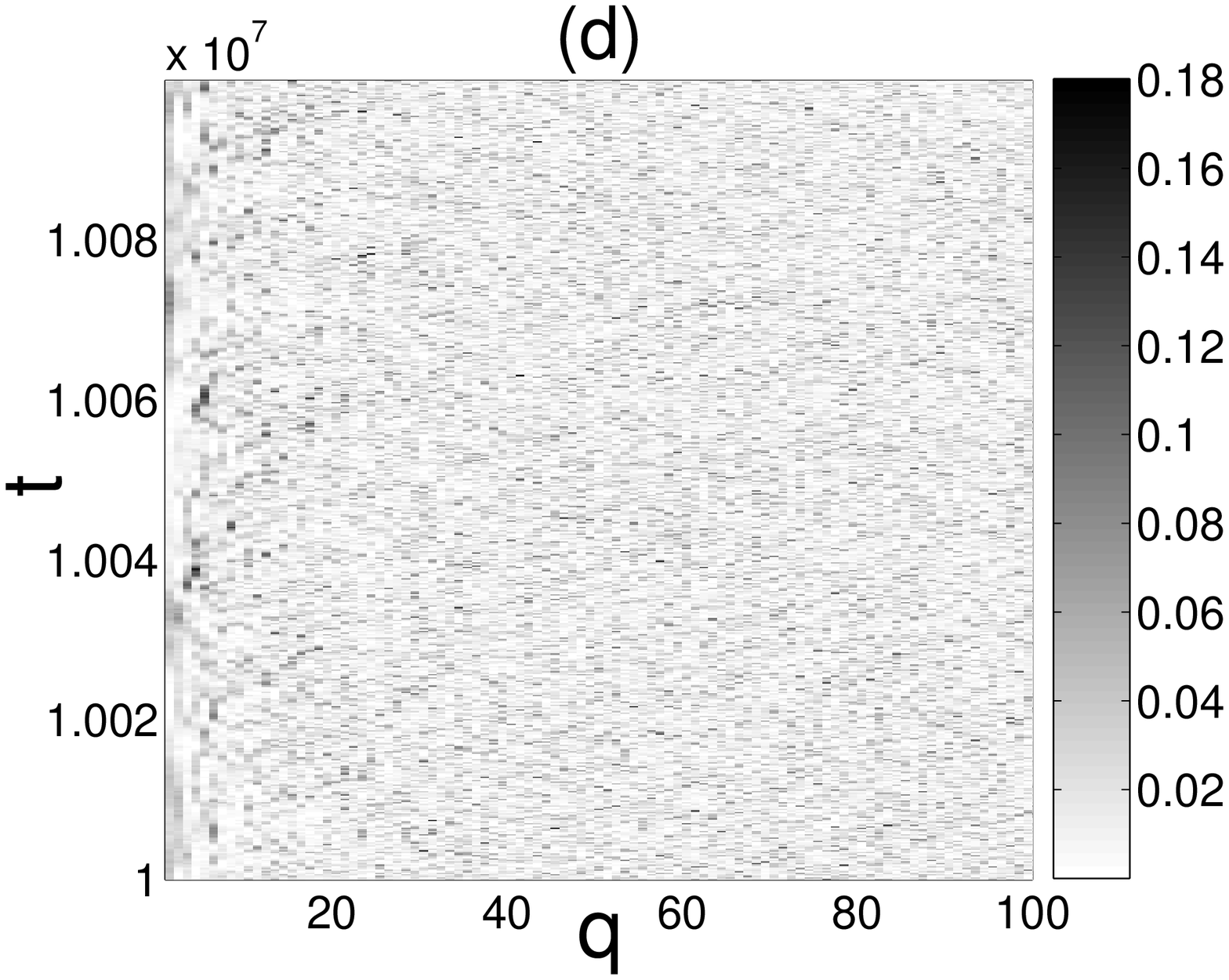}
\\
\parbox[t]{\textwidth}{\caption{Space-time plots of mode energies $E_q$
evolving from the initial localization in the third mode for
$N=100, E_{3}(0)=1.58$ and (a) $\beta=0.6, T_{tr}=10^6$, (b)
$\beta=0.6, T_{tr}=10^7$, (c) $\beta=1.25, T_{tr}=10^7$, (d)
$\beta=5.0, T_{tr}=10^7$.\label{fig8}}}
\end{figure}

In the first case we take three values of $\beta=0.6, 1.25, 5.0$,
for which energy delocalizes and essentially redistributes among the
degrees of freedom after some transition time $T_{tr}$
(Fig.\ref{fig8}). 
We note that with the chosen values of $q_0,E_0,N$ our
previous results suggest that the QB becomes unstable
at $\beta\approx 0.01$ and delocalizes at $\beta \approx 1.6$.
The figures show a time window width $10^5$ which
is much larger than the largest QB periods which are of the order of
400.

We observe that stochastic motion and
energy flow to higher modes lead the system into a possibly
long transient regime, in
which QB-like objects (finite-lifetime single-mode excitations)
are observed for all $q$ (Fig.\ref{fig8}(a,b), note
that at different $T_{tr}$ qualitatively the same picture is
observed). These objects survive for up to a hundred periods of
phonon band oscillations. 
Since the smallest chosen values for $\beta$ exceed the
QB instability threshold value, we conclude that
the QB instability is of local character in $q$-space
and does not carry a perturbed trajectory far away.
When
$\beta$ is increased, the lifetime of higher frequency QBs drops
down and they disappear in the high- and middle-frequency regions
(Fig.\ref{fig8}(c), $\beta=1.25$), and then remain observable only
in the few lowest modes (Fig.\ref{fig8}(d), $\beta=5.0$).
Note that for these values of $\beta$ we already exceed the 
delocalization threshold estimate for QBs. The observation
of surviving low-$q$ QB-like structures suggests that
the energy flow between low and high $q$ modes is sufficiently
weak, so that some excess of energy is transferred into the
large $q$ domain, allowing for long-time energy localization 
in the low $q$ domain.

\begin{figure}[p]
\includegraphics[width=0.47\textwidth]{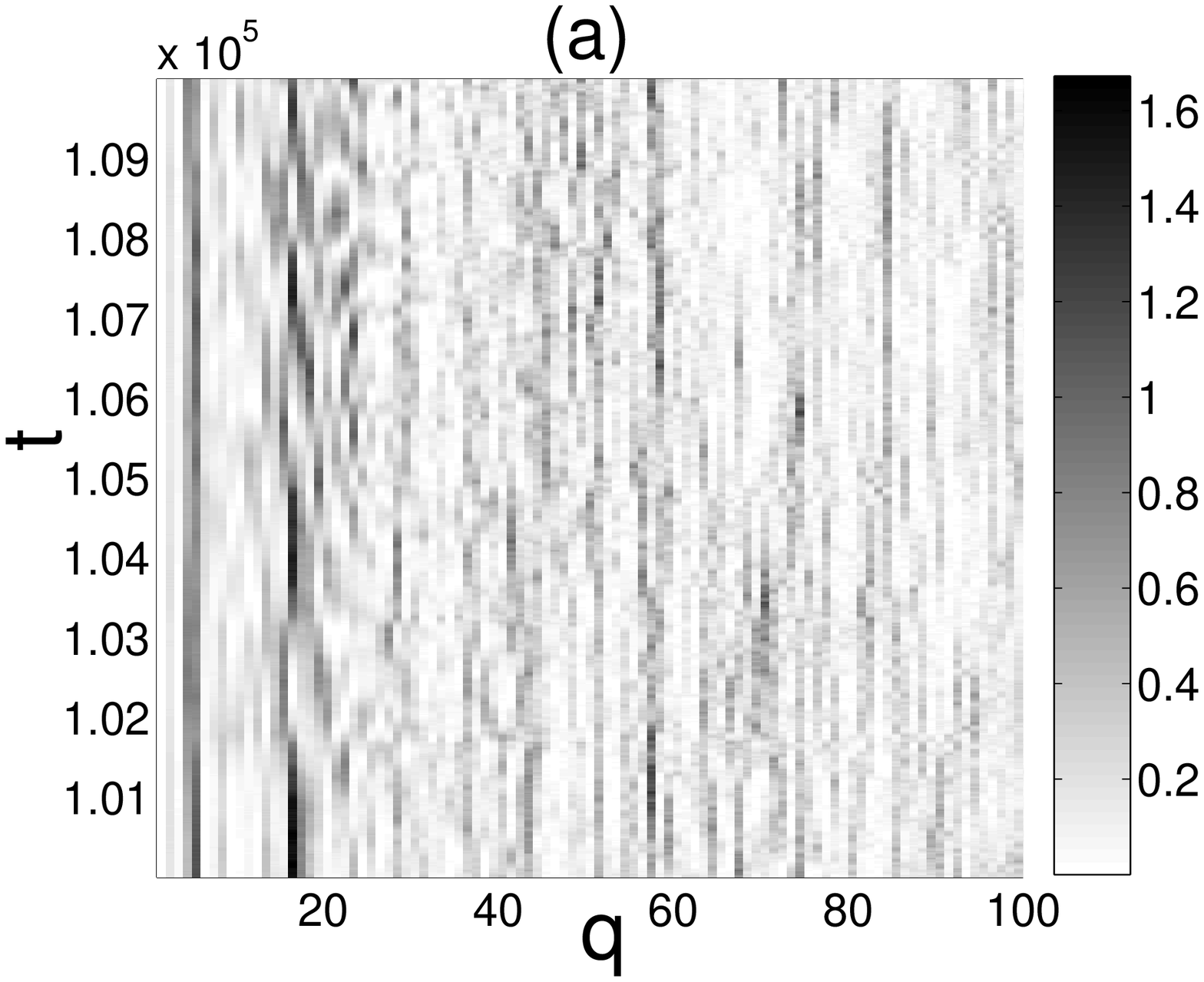}
\hfill
\includegraphics[width=0.47\textwidth]{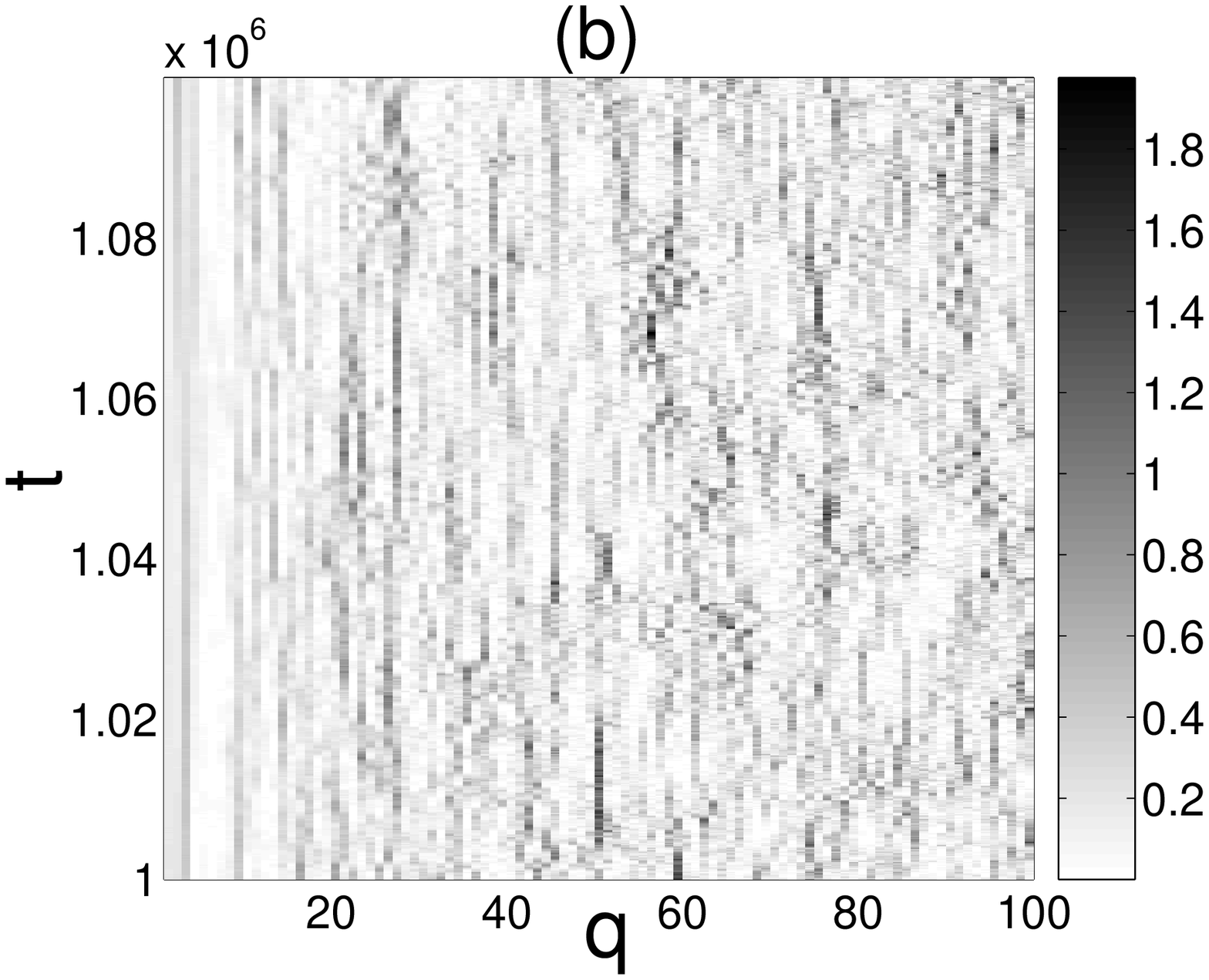}
\\
\includegraphics[width=0.47\textwidth]{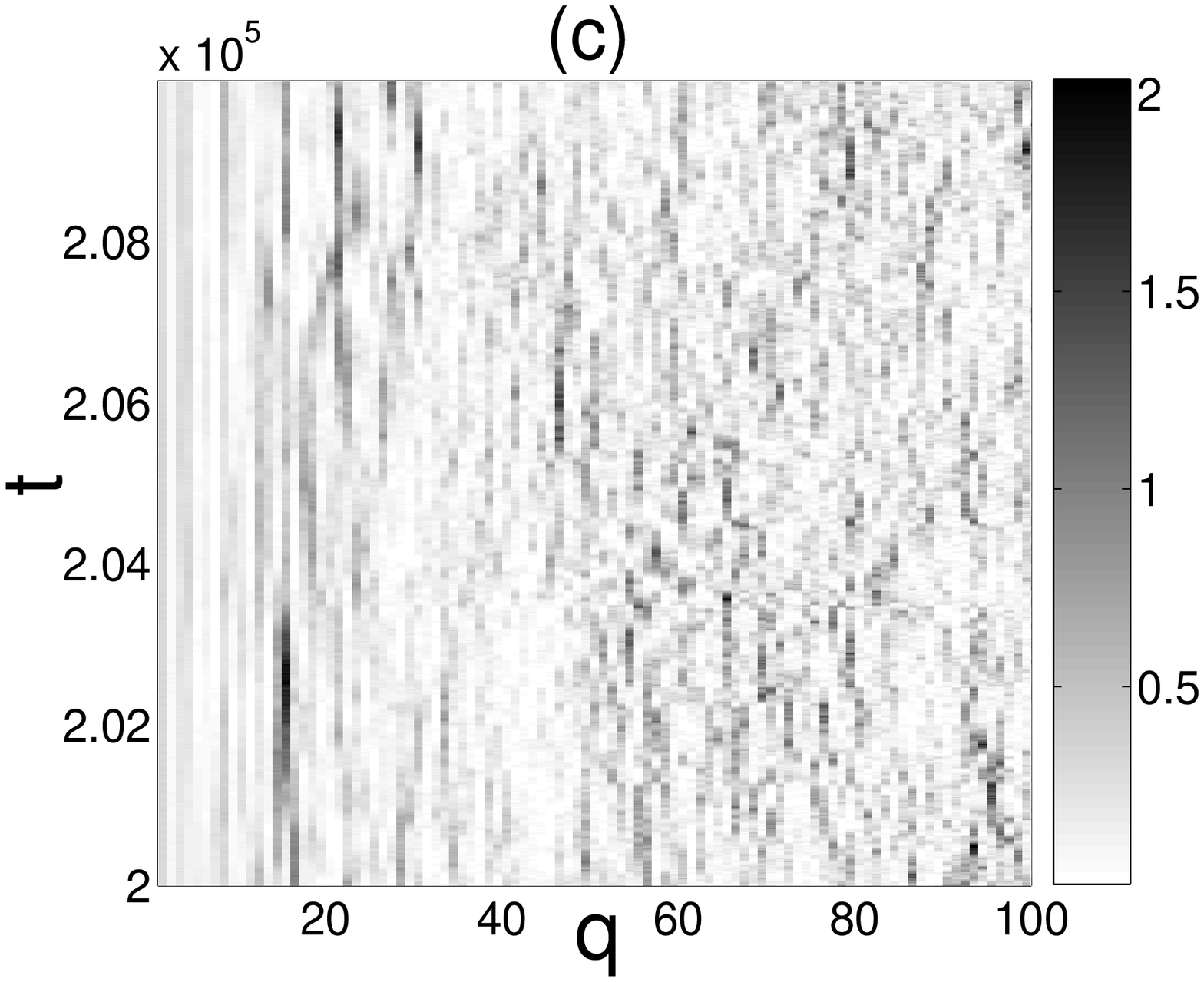}
\hfill
\includegraphics[width=0.47\textwidth]{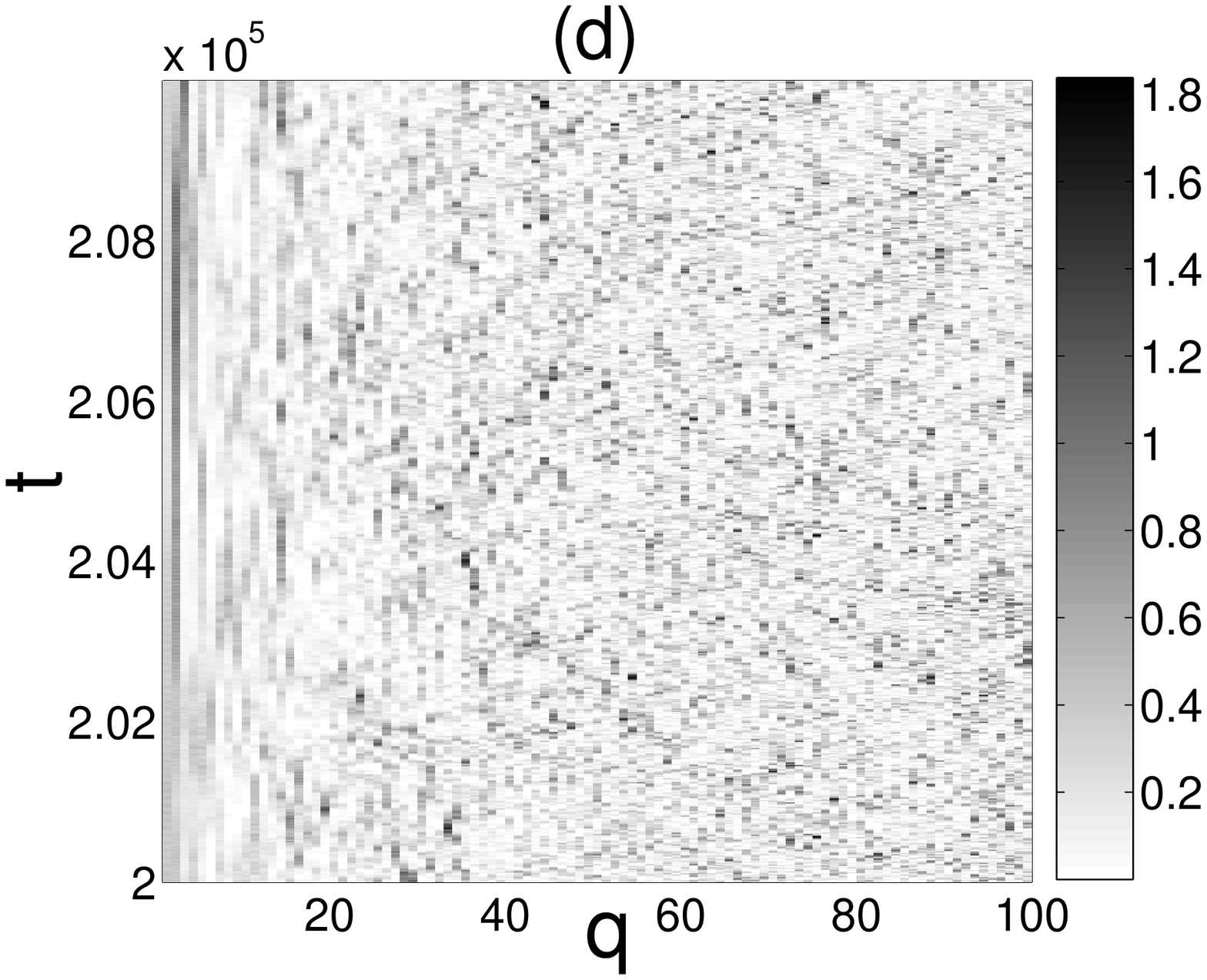}
\\
\parbox[t]{\textwidth}{\caption{Space-time plots of mode energies $E_q$
evolving from random initial conditions for $N=100, E_{tot}/N=0.2$
and (a) $\beta=0.05, T_{tr}=10^5$, (b) $\beta=0.05, T_{tr}=10^6$,
(c) $\beta=0.1, T_{tr}=2\cdot10^5$, (d) $\beta=0.4,
T_{tr}=2\cdot10^5$.\label{fig7}}}
\end{figure}

Similar effects can be observed for the second type of initial
conditions (Fig.\ref{fig7}) which mimics thermal equilibrium. 
At low $\beta$ QBs can
be observed in the whole phonon band frequency domain
(Fig.\ref{fig7}(a,b), $\beta=0.05$). As $\beta$ is increased,
high- and medium-frequency QBs disappear (Fig.\ref{fig7}(c,d),
$\beta=0.1, 0.4$). Note, that by rescaling coordinates and
momenta (and, hence, energies) the set of total energies and
nonlinear coupling strengths for both types of initial conditions
practically coincide, so one may directly compare all subfigures
from Fig.\ref{fig8} and Fig.\ref{fig7}.

\section{Conclusion}

We report on the existence of $q$-breathers as exact
time-periodic low-frequency solutions in the nonlinear FPU system.
These solutions are exponentially localized in the $q$-space of
the normal modes and preserve stability for small enough
nonlinearity. They continue from their trivial counterparts for
zero nonlinearity at finite energy. The stability threshold of QB
solutions coincides with the weak chaos threshold
in \cite{deLuca}.
The delocalization threshold estimate of QBs shows identical
scaling properties as the 
estimate of equipartition from second order nonlinear resonance overlap
\cite{Shepel}.
Persistence of exact stable QB modes is
shown to be related to the FPU observation. The FPU
trajectories computed 50 years ago are perturbations of the exact QB orbits.
Remarkably, localization in $q$-space persists even for parameters
when the QBs turn unstable. 
The QB concept - tracing time-periodic and $q$-space localized
orbits together with their stability and degree of localization -
allows to explain quantitatively and semiquantitatively many
aspects of the 50 year old FPU problem.
Moreover we show that dynamical localization in $q$-space persists
for transient processes and thermal equilibrium remarkably
well. Note that for certain cases specific symmetries allow to obtain
$q$-breathers with compact localization \cite{bushes,highsymmetry}, where they
have been interpreted as anharmonic plane wave solutions.

The concept of
QBs and their impact on the evolution of excitations in the
FPU system is expected to apply far beyond the stability threshold
of the QB solutions reported in the present work.
Generalizations to higher dimensional lattices and other
Hamiltonians are straightforward, due to the weak constraint imposed
by the nonresonance condition needed for continuation.
QBs can be also expected to contribute to peculiar dynamical features of
nonlinear lattices in thermal equilibrium, e.g.
the anomalous heat conductivity in
FPU lattices \cite{chaosfpu}.

A quantization of QBs will lead straightforwardly to quantum QBs,
which will not differ much from their classical counterparts,
due to the absence of discrete symmetries which may lead
to tunneling effects. Thus a quantum QB state will be localized
in $q$-space pretty much as a classical QB, allowing for ballistic-type
excitations in the low-$q$ domain.

Finally we remark that the QB concept is not even constrained to
an underlying lattice in real space. What we need in order to construct
a QB is a discrete spectrum of mode energies, and nonlinearity
which induces mode-mode interactions. Thus QBs can be constructed
in various nonlinear field equations on a finite spatial domain
as well, since one again obtains in the limit of small amplitudes
a discrete mode spectrum, with the only difference to a finite
lattice that the number of modes is infinite. 
\\
\\
\\
Acknowledgements
\\
\\
We thank T. Bountis, S. Denisov, 
F. Izrailev, Yu. Kosevich, A. J. Lichtenberg, P. Maniadis, 
V. Shalfeev and W. Zakrzewski
for stimulating discussions. M.I. and O.K. appreciate the warm
hospitality of the Max Planck Institute for the Physics of Complex
Systems. M.I. also acknowledges support of the "Dynasty"
foundation.

\appendix
\section{QB localization in the $\alpha$-FPU model}\label{alploc}
First, we note, that \eqref{Qzero} and \eqref{senior} are true for $k=1$,
according to \eqref{sol0}. Further, we assume it is 
true for $k=1,\dots ,n-1$ and
prove, that it is then true for $k=n$ as well.

Equation \eqref{ordern} for  the order $n-1$ can be written as
\beq[ordern1]
\Ddot Q^{(n-1)}_q+\om_q^2 Q^{(n-1)}_q=
-\om_q \mspace{-18mu} \sum_{\substack{n_{1,2}=1\\n_1+n_2=n}}^{n-1}
\sum_{l,m=1}^N \om_l \om_m B_{qlm}
Q^{(n_1-1)}_l Q^{(n_2-1)}_m
\eeq
Consider, for which mode numbers $q$ the right-hand part in \eqref{ordern1} is
non-zero. As $n_{1,2}<n$, we obtain from \eqref{Qzero_a} $Q^{(n_1-1)}_l\neq 0$
for $l=q_0, 2q_0, \dots, n_1 q_0$ only, and $Q^{(n_2-1)}_m\neq 0$ for $m=q_0,
2q_0, \dots, n_2 q_0$ only. According to the definition of $B_{qlm}$
\eqref{Bqlm} and taking into account that $nq_0\le N$, we obtain $B_{qlm}\neq
0$ for $q=l+m$ and $q=|l-m|$. Then, the maximal mode number $q$, for which the
right-hand part in \eqref{ordern1} is non-zero, equals  $q_{max}=n_1 q_0+n_2
q_0=nq_0$. This is equivalent to \eqref{Qzero_a} at $k=n$, which concludes the
induction for \eqref{Qzero}.

To prove \eqref{senior}, we write down \eqref{ordern1} for $q=nq_0$:
\begin{subequations}\label{nq0}
\beq[nq0full]
\Ddot Q^{(n-1)}_{nq_0}+\om_{nq_0}^2 Q^{(n-1)}_{nq_0}= -\om_{nq_0}
\mspace{-18mu} \sum_{\substack{n_{1,2}=1\\n_1+n_2=n}}^{n-1} \mspace{-18mu}
\om_{n_1} \om_{n_2} Q^{(n_1-1)}_{n_1 q_0} Q^{(n_2-1)}_{n_2 q_0}
\eeq
Here we have taken into account, that the only non-zero term of the sum over
$l$ and $m$ corresponds to $l=n_1 q_0$, $m=n_2 q_0$, and that $B_{nq_0,n_1 q_0,
n_2 q_0} =1$.

As $n_{1,2}<n$, the variables $Q^{(n_1-1)}_{n_1}$ and $Q^{(n_2-1)}_{n_2}$ are
expressed via \eqref{senior1}. Then \eqref{nq0full} becomes a forced harmonic
oscillator equation with a set of cosinusoidal harmonics in the right-hand
part:
\beq[nq0harm]
\Ddot Q^{(n-1)}_{nq_0}+\om_{nq_0}^2 Q^{(n-1)}_{nq_0}=
\sum_{m=0}^n C_m \cos m\om_{q_0} t
\eeq
where $C_m$ are amplitudes of the harmonics.
\end{subequations}

Omitting the free-motion part of the solution, we write down the expression for
the forced oscillations of $Q^{(n-1)}_{nq_0}$:
\beq[forced_sol]
Q^{(n-1)}_{nq_0}=\sum_{m=0}^n \frac{C_m}{\om_{nq_0}^2-m^2 \om_{q_0}^2}
\cos m\om_{q_0} t
\eeq

Note, that the resonance denominator in \eqref{forced_sol} is minimal
at $m=n$. In other words, driving frequency $n\om_{q_0}$ is the
closest to the resonance among all the harmonics. Expanding the sine function
in \eqref{omegas} into a Taylor series as $\sin x=x-x^3/6+O(x^5)$, we
approximate the mode frequencies as
\beq[omega_apx]
\om_q=4\left((q\kap)^2-\frac{1}{3}(q\kap)^4+O\left((q\kap)^6\right) \right)
\quad \mbox{where} \quad \kap=\frac{\pi}{2(N+1)} \to 0
\eeq
The mentioned minimal denominator is then expressed at $nq_0/N\to 0$ as follows:
\beq[mindenom]
\om_{nq_0}^2-n^2\om_{q_0}^2=-\frac{1}{12}n^2 (n^2-1)\om_{q_0}^4
+O\left(\left(\frac{nq_0}{N}\right)^6\right)
\eeq
At the same time the relation of this denominator to any other one with $m\neq
n$ is estimated as
\beq[reldenom]
\frac{\om_{nq_0}^2-n^2\om_{q_0}^2}{\om_{nq_0}^2-m^2\om_{q_0}^2}=
O\left(\left(\frac{nq_0}{N}\right)^2\right)\mbox{,}\quad m\neq n
\eeq

It means, that the harmonic $m=n$ dominates in the solution \eqref{forced_sol},
while all the other harmonics are small values of the order
$O\left((nq_0/N)^2\right)$ with respect to this dominating one.

Thus, \eqref{forced_sol} can be rewritten as
\beq[forced_Cn]
Q^{(n-1)}_{nq_0}=\frac{12\, C_n}{n^2 (n^2-1)\om_{q_0}^4} \cos n\om_{q_0} t
\left(1+O\left(\left(\frac{nq_0}{N}\right)^2\right)  \right)
\eeq

Inserting \eqref{senior1} (valid as $n_{1,2}<n$) into \eqref{nq0full}, we
obtain the following expression for the $n$th harmonic amplitude $C_n$ in
\eqref{nq0harm}:
\beq[Cn]
C_n=-\om_{nq_0}
\mspace{-18mu} \sum_{\substack{n_{1,2}=1\\n_1+n_2=n}}^{n-1} \mspace{-18mu}
\om_{n_1} \om_{n_2} \frac{1}{2} A_{n_1 q_0} A_{n_2 q_0}
\cdot \left(1+O\left(\left(\frac{nq_0}{N}\right)^2\right)  \right)
\eeq

Inserting \eqref{Cn} into \eqref{forced_Cn} and taking into account
\eqref{omega_apx}, we arrive at
\beq[Qn]
Q^{(n-1)}_{nq_0}=A_{nq_0} \left( \cos n\om_{q_0} t
+O\left(\left(\frac{nq_0}{N}\right)^2\right)  \right)
\eeq
where
\beq[Anq0]
A_{nq_0}=\frac{6}{n^2 (n^2-1)\om_{q_0}}
\sum_{\substack{n_{1,2}=1\\n_1+n_2=n}}^{n-1} \mspace{-18mu}
n_1 n_2 A_{n_1 q_0} A_{n_2 q_0}
\eeq

Expressing $A_{n_1 q_0}$ and $A_{n_2 q_0}$ via \eqref{senior2} (valid because
$n_{1,2}<n$) and taking into account that
$$\sum_{\substack{n_{1,2}=1\\n_1+n_2=n}}^{n-1} n_1 n_2 \equiv
\frac{1}{6}n^2 (n^2-1) \mbox{,}$$
we obtain
\beq[fin_induc]
A_{nq_0}=\frac{A_{q_0}^n}{\om_{q_0}^{n-1}}
\mbox{.}
\eeq
Equations \eqref{Qn} and \eqref{fin_induc} coincide with (\ref{senior}a,b) at
$k=n$, thus concluding the induction.

\section{QB Stability in the $\beta$-FPU model}\label{betstab}

Making a replacement \eqref{linrepl} in the equations of motion
\eqref{betmod} and keeping only linear in $\xi_q$ terms we obtain an
equation describing the dynamics of infinitesimal deviations from
the QB orbit:
\beq[eq_xi1]
\Ddot \xi_q+\om_q^2 \xi_q=-3\rho {\hat Q_{q_0}}^2(t)\om_{q_0}^2
\sum_{r=1}^N b_{qr} \xi_r +O(\rho^2,\xi_l) \mbox{,}
\eeq
where $b_{qr}=\om_q \om_r C_{q_0,q_0,q,r}$, and $O(\rho^2,\xi_l)$ denotes a
linear form of $\{\xi_l\}$ with small coefficients of the order $O(\rho^2)$.

Inserting here $\hat Q_{q_0}(t)$ from \eqref{QB},
we obtain the following Mathieu equation:
\beq[eq_xi2]
\Ddot \xi_q+\om_q^2 \xi_q=-h(1+\cos\Omega t)
\sum_{r=1}^N b_{qr} \xi_r +O(h^2,\xi_l) \mbox{,}
\eeq
where $h=3\rho E_{q_0}$, $\Omega=2\hat{\omega}$.

In the vector-matrix form it can be rewritten as follows:
\beq[Math]
\Ddot{\bxi}+\A\bxi+h(1+\cos\Omega t)\B\bxi=O(h^2\bxi),
\eeq
where $\bxi=(\xi_q)$ is a vector, $\A=(a_{qr})$ is a diagonal
matrix with elements $a_{qr}=\delta_{qr}\omega_q^2$,
$\B=(b_{qr})$ is the coupling matrix.

Further, we analyze parametric resonance in the equation \eqref{Math},
treating $h$ and $\Omega$ as independent parameters, and then recall their
dependence.

In the limit $h\to 0$ the equilibrium point $\bxi=0$ is strongly
stable for all values of $\Omega$ except for a finite number of
values $\Omega_{nkl}$ which satisfy
\beq[nkl]
\omega_k+\omega_l=n\Omega_{nkl}
\eeq
where $n$ is a natural number, and the modes
$k$ and $l$ belong to the same connected component of the
coupling graph whose connectivity is defined by the matrix \B.
Strong stability implies that this point is also stable for all hamiltonian
systems close enough to the considered one.

Each point $(\Omega_{nkl},0)$ on the plane of parameters $(\Omega,h)$ is
associated with a zone of parametric resonance. Restricting ourselves to the
case of primary resonance, which necessarily requires $n=1$ in
\eqref{nkl}, we specify the frequency $\Omega$ as
\beq[eq_delta]
\Omega=(\om_k+\om_l)(1+\delta)\mbox{,}
\eeq
where the detuning parameter $\delta$ is assumed to be the order $\delta=O(h)$.
We seek for a solution to \eqref{Math} in the form
\beq[Mathsol]
\bxi=\sum_{m=-\infty}^{+\infty} \boldsymbol{f}^m
e^{(i\Tilde{\omega}_1+z+im\Omega)t}+\mbox{c.c.}
\eeq
where
$\Tilde{\omega}_1=\omega_k(1+\delta)$,
$\boldsymbol{f}^m=(f_q^m)$ are unknown complex vector amplitudes, and $z$
is a small unknown complex number. We make an assumption $z=O(h)$
which is confirmed further in the course of the calculation.

Inserting \eqref{Mathsol} into \eqref{Math} we obtain a system of algebraic
equations for the amplitudes $f_q^m$:
\begin{multline}\label{eq_fqm}
\left[2iz(\Tilde{\omega}_1+m\Omega)-(\Tilde{\omega}_1+m\Omega)^2+\om_q^2 \right]
f_q^m+\\
+(h\B \f^m)_q+
\left(\frac{h}{2} \B (\f^{m+1}+\f^{m-1})\right)_q = O(h^2|\bxi|)
\end{multline}

Note, that if the coefficient in square brackets in \eqref{eq_fqm} is not a
small value of the order $O(h)$, then the corresponding amplitude $f_q^m$ is
itself a small value of the order $O(h|\bxi|)$. Let us find out, for which
values of the indices $m$ and $q$ the mentioned coefficient is small. As we
assume $z=O(h)$, the difference $-(\Tilde{\omega}_1+m\Omega)^2+\om_q^2$ needs
to be a small value. It will be so, if the absolute value
$|\Tilde{\omega}_1+m\Omega|$ is close to one of the eigenfrequencies $\om_q$.
According to the definition of $\Tilde{\omega}_1$ and the expression for
$\Omega$ \eqref{eq_delta}, this implies $|(m+1)\om_k+m\om_l|=\om_q$.
Generically, due to the incommensurate eigenfrequencies spectrum, this
condition is only fulfilled for $m=0$, $q=k$ or $m=-1$, $q=l$.

It follows, that all amplitudes $f_q^m$ except for $f^0_k$ and $f^{-1}_l$ are
small values of the order $O(h|\bxi|)$. Then we are able to write down a closed
system for $f^0_k$ and $f^{-1}_l$ accurate to $O(h^2|\bxi|)$:
\begin{subequations}\label{eq_fqmred}
\begin{align}
(2iz\Tilde{\omega}_1+2\om_k^2(h-\delta))f^0_k
 + \frac{h}{2}b_{kl}f^{-1}_l&=O(h^2|\bxi|)
\\
\frac{h}{2}b_{lk}f^0_k
 + (-2iz\Tilde{\omega}_2+2\om_l^2(h-\delta))f^{-1}_l&=O(h^2|\bxi|)
\end{align}
\end{subequations}
where $\Tilde{\omega}_2=\Omega-\Tilde{\omega}_1=\om_l(1+\delta)$. Note, that
for a primary resonance the coupling coefficient $b_{kl}$ must be non-zero,
which means that the mode oscillators $k$ and $l$ are directly coupled.

A nontrivial solution to this system exists if the determinant of the left-hand
part (with an error $O(h^2)$ allowed for in each element) equals zero.
From this condition we derive
\beq[eq_z1]
z_{1,2}=-i\frac{(h-\delta)(\om_l-\om_k)}{2} \pm
\frac{1}{2} \sqrt{\frac{h^2}{4}\om_k \om_l-(h-\delta)^2(\om_k+\om_l)^2+O(h^3)}
+O(h^2)
\eeq

In the initial problem both $h$ and $\Omega$ depend on the nonlinearity
magnitude $\beta E_{q_0}$. This dependence defines a line starting from the
point $(2\omega_{m_0},0)$ on the $(\Omega,h)$ plane. The intersections of this
line with the resonance zones are the regions of the QB orbit instability.

The nearest primary resonance corresponds to $k=m_0-1$, $l=m_0+1$, $n=1$. In
this case we can derive a simpler expression for $z_{1,2}$ in the vicinity of
the bifurcation point (near the edge of the resonance zone) if we let
$N\to\infty$ at the same time assuming $q_0/(N+1)=const$ which means
$\om_{q_0}=const$.

From \eqref{omega_apx} we obtain
\begin{subequations}\label{eq_prim1}
\begin{align}
\om_{q_0+1}-\om_{q_0-1}&= 2\cos q_0\kap \cdot(1+O(\kap^2))
\\
\om_{q_0+1}\om_{q_0-1} &= \om_{q_0}^2 \cdot(1+O(\kap^2))
\\
\om_{q_0+1}+\om_{q_0-1}&= 2\om_{q_0} \cdot(1+O(\kap^2))
\end{align}
\end{subequations}

Then we express $\delta$ from \eqref{eq_delta}:
\beq[eq_delta2]
\delta=\frac{\Omega}{\om_{q_0+1}+\om_{q_0-1}}-1=\frac{3}{4}h+\frac{1}{2}\kap^2
+O(\kap^4)
\eeq

Inserting (\ref{eq_prim1}a-c) and \eqref{eq_delta2} into \eqref{eq_z1} and
taking into account that if we are close to the bifurcation point (the
expression under the square root in \eqref{eq_z1} is close to zero) then
$h=O(\kap^2)$, we find
\beq[eq_z2]
z_{1,2}=\pm \frac{1}{2} \om_{q_0} \kap^2 \sqrt{R-1+O(\kap^2)}+i\cdot O(\kap^3)
\eeq
where $R=h/\kap^2=6\beta E_{q_0}(N+1)/\pi^2$.

According to \eqref{Mathsol}, the absolute values of the Floquet multipliers
involved in the resonance are calculated as
\beq\label{mults0}
|\lambda_{j_1 j_2}|=\exp(\frac{2\pi\mbox{Re} z_{1,2}}{\Omega}) \mbox{,}
\eeq
which finally leads to \eqref{mults}.


\begin{thebibliography}{99}
\bibitem{symmanifolds}
Note that we do not demand specific symmetries leading to
low-dimensional invariant manifolds in phase space, in contrast
to the discussion in \cite{bushes,highsymmetry}.
Consequently, our QB solutions do not have to be embedded in such
subspaces.
\bibitem{corrections}
The subsequent bifurcations of the same type are associated with
next primary resonances of the type $k=q_0-m$, $l=q_0+m$, $n=1$,
$m=2,3,\dots$. The corresponding bifurcation points are located
at $R=m^2+O(1/N^2)$. Note, that higher-order resonances at
certain values of $q_0$ and $N$ may result in appearance of
narrow instability regions  within the region $0<R<1$.
\end{thebibliography}
\end{document}